\documentclass[12pt, a4paper]{article}
\usepackage[dvipdfmx]{graphicx}
\usepackage{amssymb}
\usepackage{amsmath}
\usepackage{bm}
\usepackage{xcolor}
\usepackage{cite}
\usepackage{slashed}
\usepackage{subfigure}
\usepackage{epstopdf}            
\usepackage{epsfig}
\usepackage{here}
\usepackage{comment}
            
\renewcommand{\thefootnote}{\fnsymbol{footnote}}
\numberwithin{equation}{section} % Eq.(Sec.eq.)
\def\beq#1\eeq{\begin{align}#1\end{align}}

\newcommand{\eg}{{\em e.g.}}
\newcommand{\ie}{{\em i.e.}}
\newcommand{\non}{\nonumber \\ }

\newcommand{\emu}{e\mu}

\RequirePackage{xspace}
\def\Bbar    {\kern 0.18em\overline{\kern -0.18em B}{}\xspace}

\def\Kbar    {\kern 0.18em\overline{\kern -0.18em K}{}\xspace}

\usepackage{caption}
\definecolor{BlueViolet}{rgb}{0.2, 0.00, 0.7}
\definecolor{Blue}{rgb}{0.15, 0.00, 0.9}
\usepackage[
colorlinks=true, linkcolor=Blue, citecolor=Blue, urlcolor=BlueViolet]{hyperref} 
\usepackage[utf8]{inputenc}
\begin{document}

\begin{titlepage}
\begin{center}

\hfill KEK--TH--2189

\vskip 1.in

{\LARGE \bf
Probing $\boldsymbol{e\mu}$ flavor-violating ALP at Belle II}
 
\vskip .4in

{\large 
Motoi Endo$^{\rm (a,b)}$,
Syuhei Iguro$^{\rm (c)}$,
and 
Teppei Kitahara$^{\rm (d,e,f)}$
}
\vskip 0.25in

{\small 
$^{\rm (a)}${\it 
KEK Theory Center, IPNS, KEK, Tsukuba 305-0801, Japan}
\vskip 0.1in

$^{\rm (b)}${\it 
The Graduate University of Advanced Studies (Sokendai), Tsukuba 305-0801, Japan}
\vskip 0.1in

$^{\rm (c)}${\it 
Department of Physics,
Nagoya University, Nagoya 464-8602, Japan}
\vskip 0.1in

$^{\rm (d)}${\it 
Physics Department, Technion--Israel Institute of Technology, Haifa 3200003, Israel}
\vskip 0.1in

$^{\rm (e)}${\it 
Institute for Advanced Research, Nagoya University, Nagoya 464-8601, Japan}
\vskip 0.1in

$^{\rm (f)}${\it 
Kobayashi-Maskawa Institute for the Origin of Particles and the Universe, 
Nagoya University,  Nagoya 464-8602, Japan}
}
\vskip 1.in

\begin{abstract}
\noindent
Recently, 
it was pointed out that 
the electron and muon $g-2$ discrepancies can be explained simultaneously 
by a single flavor-violating axion-like particle (ALP).
We show that 
the parameter regions favored by the muon $g-2$ are already excluded by the muonium-antimuonium oscillation bound.
In contrast, those for the electron $g-2$ can be consistent with this bound
when the ALP is heavier than $1.5$\,GeV.
We propose to search for a signature of the same-sign and same-flavor lepton pairs and the forward-backward muon asymmetry to test the model at the Belle II experiment.
\end{abstract}
\end{center}
\noindent ~~~~~~~\,\textsc{Keywords:}
Electron $g-2$, Muon $g-2$, Axion-like particle, Belle II experiment 

\end{titlepage}

\setcounter{page}{1}
\renewcommand{\thefootnote}{\#\arabic{footnote}}
\setcounter{footnote}{0}

%%%%%%%%%%%%%%%%%%%%%%%%%%%%%%%%%%%%
\section{Introduction}
\label{sec:introduction}
%%%%%%%%%%%%%%%%%%%%%%%%%%%%%%%%%%%%

A long-standing discrepancy between the experimental result and the Standard Model (SM) prediction 
of the anomalous magnetic dipole moment of the muon, $(g-2)_{\mu}$,
may be a signal of new physics beyond the SM (BSM).
Experimentally, it has been determined precisely by the Brookhaven E821 experiment \cite{Bennett:2002jb,Bennett:2004pv,Bennett:2006fi}. 
(See Ref.~\cite{Endo:2020mqz} for the latest result updating the value of the muon-to-proton magnetic ratio.)
By virtue of dedicated theoretical efforts 
\cite{Aoyama:2012wk,Aoyama:2017uqe,Volkov:2019phy,Knecht:2002hr,Czarnecki:2002nt,Gnendiger:2013pva,Ishikawa:2018rlv,Davier:2019can,Keshavarzi:2019abf,Prades:2009tw,Blum:2019ugy},
the latest result of $a_\mu \equiv (g - 2)_\mu /2$ is obtained as
\begin{align}
\Delta a_\mu=
a_{\mu}^{\rm exp} - a_{\mu}^{\rm SM} =
\begin{cases}
\left( 27.8 \pm 7.4\right) \times 10^{-10} & \text{\cite{Keshavarzi:2019abf}}\,,\\
\left( 26.1 \pm 7.9\right) \times 10^{-10} & \text{\cite{Davier:2019can}}\,,
\end{cases}
\label{eq:Deltaamu}
\end{align}
depending particularly on the determination of the hadronic vacuum polarization.
They correspond to $3.8$ and $3.3\,\sigma$ discrepancies, respectively. 
Currently,
a new measurement of $a_\mu$ is in progress at the Fermilab \cite{Grange:2015fou, Keshavarzi:2019bjn}, aiming to reduce the experimental uncertainty by a factor of $4$ compared to the Brookhaven result.
Another experiment is planned at the J-PARC~\cite{Mibe:2011zz, Abe:2019thb}, where distinct methods are employed to measure $a_\mu$.

Recently, a new candidate for the BSM signal has been reported in the electron $g-2$, $(g-2)_e$.
The fine-structure constant was determined precisely
by measuring the caesium mass  
\cite{Parker:2018vye}. 
Consequently, the theoretical uncertainty of $a_{e}$ was reduced by a factor of $3$ compared to the previous result based on the measurement of the rubidium mass \cite{Bouchendira:2010es}.
It revealed that the experimental result \cite{Hanneke:2008tm} deviates from the SM prediction \cite{Aoyama:2017uqe} (cf.~Ref.~\cite{Volkov:2019phy}) as
\begin{align}
\Delta a_e=a_{e}^{\rm exp} - a_{e}^{\rm SM} = (-0.88 \pm0.36) \times10^{-12}\,,
\label{eq:Deltaae}
\end{align}
corresponding to a $2.5\,\sigma$ discrepancy. 

It is noticed that the experimental value of $a_{e}$ is {\it smaller} than the SM prediction.
This is in contrast to the result of $a_\mu$, where the experimental value is {\it larger} than the SM one. 
Although it is not difficult to explain either one of the $g-2$ anomalies by BSM models,
it is challenging to explain both of them simultaneously by a single model because of this sign difference.
Furthermore, magnitude of the electron $g-2$ discrepancy is large:
If $\Delta a_\mu$ is explained by a BSM model, the electron $g-2$ is expected to receive a contribution from this model of $\sim m_e^2/m_\mu^2 \times \Delta a_\mu =\mathcal{O}(10^{-13})$, because the BSM contributions are proportional to the lepton mass squared in a wide class of models. 
It is obvious that $\mathcal{O}(10^{-13})$ is too small to saturate the discrepancy in Eq.~\eqref{eq:Deltaae}.
Therefore, an additional mechanism, which breaks the lepton mass squared scaling, is required to enhance the contribution to the electron $g-2$~\cite{Giudice:2012ms}.

In literature, many BSM models have been proposed to explain the anomalies simultaneously  \cite{Crivellin:2018qmi,Liu:2018xkx,Endo:2019bcj,Badziak:2019gaf,CarcamoHernandez:2019ydc,Davoudiasl:2018fbb,Dutta:2018fge,Han:2018znu,Abdullah:2019ofw,Bauer:2019gfk,Cornella:2019uxs}.
In particular, an axion-like particle (ALP) with lepton-flavor violating (LFV) interactions was introduced recently \cite{Bauer:2019gfk,Cornella:2019uxs}.\footnote{%
To explain
 the mass hierarchy and mixing matrices of the matter fermions,
flavor symmetries (horizontal symmetries) are often introduced.
Flavor-violating ALPs have been originally proposed 
as a pseudo-Nambu-Goldstone boson connected with the spontaneous flavor symmetries breaking at a high scale \cite{Davidson:1981zd,Reiss:1982sq,Gelmini:1982zz,Wilczek:1982rv,Anselm:1985bp,Feng:1997tn}, and are called flavons or familons.
}
Although the ALP is likely to have small couplings, it is expected to be light, and thus, its contribution to the $g-2$ can be large.
The sign difference between $\Delta a_e$ and $\Delta a_\mu$ is accommodated if axial-vector couplings are larger than vector ones.
Besides, the LFV couplings of the ALP enable us to enhance the contribution to the electron $g-2$ due to a chirality enhancement.

In this paper, we study 
 a unique LFV ALP model which can accommodate the electron and muon $g-2$ anomalies.\footnote{%
See also Refs.~\cite{Cornella:2019uxs,MartinCamalich:2020dfe} for phenomenology of general flavor-violating ALP models.
} 
Since the ALP is a real scalar field, its couplings to the electron and muon induce the transition of the muonium into antimuonium.
It will be shown that the parameter regions which explain both of the $g-2$ anomalies are already excluded by this process.
In particular, the ALP contributions to the muon $g-2$ are tightly constrained. 
In contrast, those to the electron $g-2$ can be consistent with it. 
We will study future prospects of the Belle II experiment to probe such parameter regions.
In this paper,
the following signatures are evaluated;
(1) a production of the same-sign and same-flavor lepton pairs, $e^+ e^- \to  \mu^{\pm} \mu^{\pm}e^{\mp} e^{\mp}$ via an on-shell ALP production, 
and (2) a forward-backward (FB) asymmetry in muon pair production, $e^+ e^- \to \mu^+ \mu^-$. 

This paper is organized as follows. 
In Sec.~\ref{sec:ALP},
we briefly introduce a flavor-violating ALP model.
In Sec.~\ref{sec:gmin2},
the ALP contributions to the electron and muon $g-2$ are explained, 
and 
the bound of the muonium-antimuonium oscillation is examined in Sec.~\ref{sec:muonium}.
Future prospects at the Belle II experiment are investigated 
in Sec.~\ref{sec:BelleII}:
In Sec.~\ref{sec:onshell}, a signature of the on-shell production of the ALP is studied, and the FB asymmetry of the muon is discussed in Sec.~\ref{sec:AFB}.
Finally, Sec.~\ref{sec:conclusion} is devoted to conclusions and discussion.
In Appendix~\ref{App:muonium}, the ALP contribution to the transition probability of the muonium into antimuonium is derived, and an analytic formula of the FB asymmetry is provided in Appendix~\ref{App:AFB}.

%%%%%%%%%%%%%%%%%%%%%%%%%%%%%%%%%%%%
\section{Flavor-violating ALP model}
\label{sec:ALP}
%%%%%%%%%%%%%%%%%%%%%%%%%%%%%%%%%%%%

Let us introduce ALP ($a$), which is a real (pseudo-) scalar field, with flavor-violating interactions. 
Since such an ALP is regarded as a pseudo-Nambu-Goldstone boson of a broken symmetry, 
the mass $m_a$ becomes naturally small compared to the broken scale $\Lambda$.
The low-energy effective Lagrangian is obtained as
\cite{Georgi:1986df,Bauer:2019gfk,Cornella:2019uxs}
\begin{align}
{\cal L}_{\rm eff} &= \frac{1}{2} \left[\left(\partial_{\mu} a\right)^2 - m_a^2 a^2\right]
\notag \\ &\quad
- \frac{\partial_\mu a}{\Lambda}\sum_{i,j}\bar{f_i}\gamma^\mu (v_{ij}-a_{ij}\gamma_5)f_j 
+ c_{\gamma \gamma}^{\rm eff} \frac{\alpha}{4 \pi} \frac{a}{\Lambda} F_{\mu \nu}\tilde{F}^{\mu \nu}
+ c_{g g}^{\rm eff} \frac{\alpha_s}{4 \pi} \frac{a}{\Lambda} G_{\mu \nu}\tilde{G}^{\mu \nu}\,,
\label{eq:Leff}
\end{align}
up to dimension-five operators.
Here, $v_{ij}$, $a_{ij}$, $c_{\gamma \gamma}^{\rm eff}$ and $c_{g g}^{\rm eff}$ are dimensionless parameters, which depend on details of the UV models. 
In particular, $v_{ij}$ and $a_{ij}$ are Hermitian matrices with flavor indices $i,j$. 
Here, $f_{i}$ denotes a matter fermion including the lepton in the $i$-th generation in the mass basis. 

For the flavor-conserving interaction ($i=j$), it is noticed that contributions of $v_{ii}$ vanish automatically and only the pseudo-scalar term is left.
This is obvious if we consider a case when the ALP couples to on-shell fermions.
Then, the $a \bar f_i f_j $ interaction is rewritten as
\begin{align}
{\cal L}_{\rm eff}&= i\frac{a}{\Lambda} 
\sum_{i,j}\bar{f_i}\left[(m_i -m_j )v_{ij}-(m_i+m_j)a_{ij} \gamma_5 \right]f_j\,,
\label{eq:LeffOnShell}
\end{align}
by using the equation of motion, where $m_i$ is the mass of $f_i$. 
For $i=j$, the coefficient of $v_{ii}$ is found to be zero. 
Note that $c_{\gamma \gamma}^{\rm eff}$ and $c_{gg}^{\rm eff}$ are induced generally by fermion one-loop diagrams as $c_{\gamma \gamma}^{\rm eff},\,c_{gg}^{\rm eff} = \mathcal{O}(a_{ii})$, even if they are suppressed in a high energy scale \cite{Bauer:2017ris,Bauer:2019gfk}.

In this paper, the ALP interactions are assumed to satisfy
\beq
 |v_{\emu}|, \,|a_{\emu}|
 \gg |v_{e\tau}|, \,|a_{e\tau}|, \,|v_{\mu\tau}|, \,|a_{\mu\tau}| 
 \gg |a_{ii}|,\, |c_{\gamma \gamma}^{\rm eff}|, \,|c_{gg}^{\rm eff}|\,,
\eeq
to evade severe constraints from LFV observables.\footnote{%
We ignore ALP interactions with quarks because they are irrelevant for the lepton $g-2$.
The interactions with neutrinos are also irrelevant in the following analysis.}
The second inequality is imposed because $v_{e\mu}$, $a_{e\mu}$ can induce LFV decays of muons such as $\mu\to e\gamma$, $\pi\to e\mu$, $\mu \to eee $, and $\mu\to e+{\rm invisible}$ when combined with $a_{ii}$, $c_{\gamma \gamma}^{\rm eff}$ or $c_{gg}^{\rm eff}$ \cite{Bauer:2019gfk,Cornella:2019uxs,Dev:2017ftk}.
On the other hand, the first condition is required because other types of LFV decays such as $\tau \to \mu \gamma$, $\tau \to e \gamma$, $\tau\to \mu \mu e$, and $\tau\to \mu e e$ can be generated by combining $v_{e\mu}$, $a_{e\mu}$ with those including the tau lepton \cite{Cornella:2019uxs,Dev:2017ftk}.
Such a hierarchy between the ALP interactions 
could be obtained from $Z_4$ lepton flavor symmetry (cf.~Ref.~\cite{Abe:2019bkf}).

From Eq.~\eqref{eq:Leff}, it is noticed that $v_{ij}$ and $a_{ij}$ appear in associated with $\Lambda$ in the expressions of ALP contributions to observables.
It is convenient to define dimensionless scalar coupling $\left(y_{V}\right)_{\emu}$ and pseudo-scalar coupling $\left(y_{A}\right)_{\emu}$ as\footnote{%
The couplings $y_{V}$ and $y_{A}$ are related to those in Ref.~\cite{Bauer:2019gfk} as
\beq
\left(y_{V}\right)_{\emu} =
 i \frac{m_{\mu}}{f} \frac{(k_e + k_E)_{\emu}}{2}\,,
 \quad 
 \left(y_{A}\right)_{\emu} =
 - i \frac{m_{\mu}}{f} \frac{(k_e - k_E)_{\emu}}{2}\,.
 \notag
\eeq
}
\beq
\left(y_{V}\right)_{\emu} = - i \frac{m_{\mu}}{\Lambda} v_{\emu}\,,
\quad \left(y_{A}\right)_{\emu} = - i \frac{m_{\mu}}{\Lambda} a_{\emu}\,.
\eeq
For instance, by neglecting the electron mass, the interactions \eqref{eq:LeffOnShell} are expressed as
\begin{align}
\mathcal{L}_{\rm eff} &\simeq 
- i\frac{m_\mu }{\Lambda} a \bar{e}  \left( v_{e\mu} + a_{e\mu} \gamma_5 \right) \mu 
+ i\frac{m_\mu }{\Lambda} a \bar{\mu}\left( v_{ e\mu}^\ast - a_{e\mu}^\ast \gamma_5 \right) e\nonumber\\
&= 
\left(y_{V}\right)_{\emu} a \bar e \mu
+ \left(y_{A}\right)_{\emu} a \bar e \gamma_5 \mu + \textrm{H.c.}\,.
\label{eq:LeffYukawa}
\end{align}
Note that $v_{\mu e} = v_{e \mu}^{\ast}$ and $a_{\mu e} = a_{e \mu}^{\ast}$ by  Hermiticity.
Therefore, the ALP contributions are represented by the following parameters,
\begin{align}
m_a,\quad
\left(y_{V}\right)_{\emu},\quad
\left(y_{A}\right)_{\emu}.
\end{align}
Note that the partial wave unitarity sets the upper bound \cite{Cornella:2019uxs}:
\beq
\left|\left(y_{V,A}\right)_{\emu} \right| \lesssim 2 \sqrt{ \frac{2 \pi}{3}}\simeq2.9\,.
\eeq

%%%%%%%%%%%%%%%%%%%%%%%%%%%%%%%%%%%%
\section{Lepton \texorpdfstring{\mbox{\boldmath$g-2$}}{g-2}}
\label{sec:gmin2}
%%%%%%%%%%%%%%%%%%%%%%%%%%%%%%%%%%%%

%%%%%%%%%%%%%%%%%%%
\begin{figure}[t]
  \begin{center}
    \includegraphics[width=8cm]{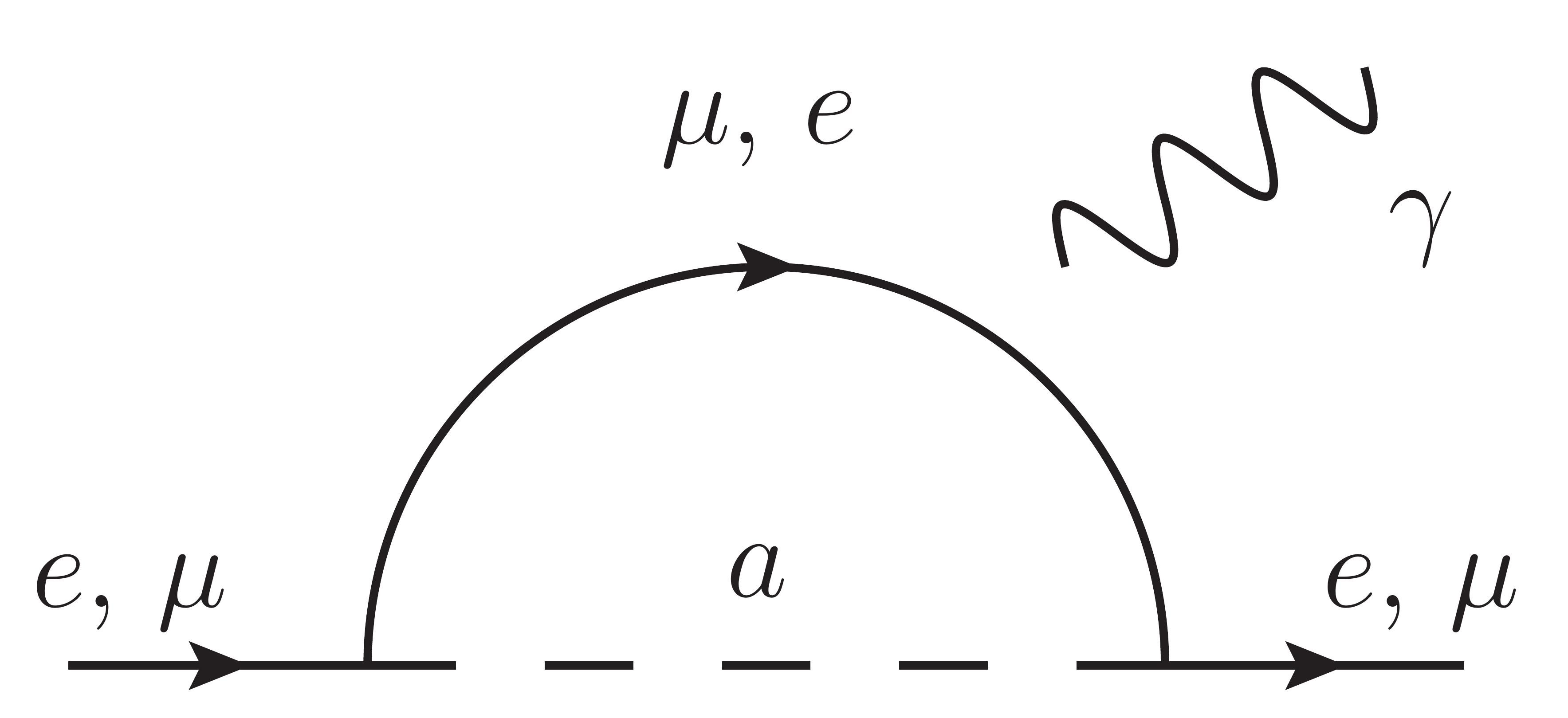}
        \caption{
        Feynman diagram for the ALP contributions to the electron and muon $g-2$. 
        Contributions from $a\gamma \gamma$ and flavor-diagonal ALP interactions are assumed to be suppressed.
            }
    \label{fig:diagram}
  \end{center}
\end{figure}
%%%%%%%%%%%%%%%%%%%%%

The ALP contributions to the lepton $g-2$ are generated at the one-loop level as displayed in Fig.~\ref{fig:diagram}.
From the effective Lagrangian in Eq.~\eqref{eq:Leff}, we obtain the results as,\footnote{%
In this paper, we focus on the flavor-violating ALP contributions.
However, it was argued that the electron and muon $g-2$ discrepancies can also  be explained simultaneously by large flavor-conserving ALP couplings as
$|c_{\gamma \gamma}^{\rm eff}|\gg |a_{\mu \mu}| \gg |v_{ij}|,|a_{ij}|_{i\neq j}$ \cite{Bauer:2019gfk,Cornella:2019uxs}.
}
%%%%%%%%%%%%%%%%%%%%%
\begin{align}
 a_e^{\rm ALP}
&=- \frac{1}{16\pi^2 }\frac{m_e }{m_{\mu}}\left\{ |\left(y_{A}\right)_{\emu}|^2- |\left(y_{V}\right)_{\emu}|^2\right\}\left[\frac{2x_{a}^2 {\rm log}\,x_{a}}{(x_{a}-1)^3}+\frac{1-3x_{a}}{(x_{a}-1)^2} \right] \,,\label{eq:delta_ae}\\
a_\mu^{\rm ALP}
&=\frac{1}{16\pi^2} \left\{|\left(y_{A}\right)_{\emu}|^2+|\left(y_{V}\right)_{\emu}|^2\right\}\left[2x_{a}^2 {\rm log}\left(\frac{x_{a}}{x_{a}-1}\right)-1-2x_{a} \right]\,,\label{eq:delta_amu}
\end{align}
%%%%%%%%%%%%%%%%%%%%%
for $m_e^2 \ll m_{\mu}^2, m_a^2$, where $x_{a}=m_a^2/m_\mu^2$. 
These results are consistent with those in Refs.~\cite{Bauer:2019gfk, Cornella:2019uxs}.\footnote{%
For a model with the flavor-violating Yukawa interactions~\eqref{eq:LeffYukawa},
the muon $g-2$ was studied in Ref.~\cite{Nie:1998dg} with providing the exact formula, and its relation to the electron $g-2$ was discussed in Ref.~\cite{Galon:2016bka}.
}
Although $a_\mu^{\rm ALP}$ seems to diverge at $x_a = 1$, this is because the approximation becomes invalid; at $x_a = 1$, the result becomes
\begin{align}
a_\mu^{\rm ALP}|_{x_a = 1}
&=\frac{1}{16\pi^2} 
\left\{|\left(y_{A}\right)_{\emu}|^2+|\left(y_{V}\right)_{\emu}|^2\right\}
\left[{\rm log}\left(\frac{m_{\mu}^2}{m_e^2}\right)-3 \right]\,,
\end{align}
for $m_e^2 \ll m_{\mu}^2$.
As pointed out in Ref.~\cite{Bauer:2019gfk}, the sign of $a_e^{\rm ALP}$ can  be opposite to that of $a_\mu^{\rm ALP}$ particularly when $|\left(y_{A}\right)_{\emu}| > |\left(y_{V}\right)_{\emu}|$ is satisfied.
In fact, the loop function in $a_e^{\rm ALP}$ is positive for any $x_a$, while the  function in $a_{\mu}^{\rm ALP}$ is positive when $x_a \gtrsim 0.9$.

For $|\left(y_{A}\right)_{\emu}| \gg  |\left(y_{V}\right)_{\emu}|$ with $x_a \sim 1$, the ALP contributions are scaled as
\beq
a_e^{\rm ALP} \sim - \frac{m_e}{m_{\mu}} a_{\mu}^{\rm ALP}\,.
\eeq
This result is understood as follows: 
the definition of the anomalous magnetic moment $a_\ell$ is normalized by the lepton mass $m_\ell$ and requires a chirality flip for the lepton. In the ALP contributions (see  Fig.~\ref{fig:diagram}), the latter for the electron $g-2$ is caused by the intermediate muon, while it is provided by the external muon for the muon $g-2$. 
Thus, the ratio is scaled by the single power of the lepton mass.
Then, if the ALP contribution to the muon $g-2$ is comparable to $\Delta a_{\mu}$, the contribution to the electron $g-2$ becomes as large as $\mathcal{O}(10^{-11})$, which overshoots $\Delta a_e$ in Eq.~\eqref{eq:Deltaae} by an order of magnitude.
Therefore, 
to explain the anomalies simultaneously,
a parameter tuning is necessary between $|\left(y_{V}\right)_{\emu}| $ and $ |\left(y_{A}\right)_{\emu}|$ at an order of $1$--$10\%$ levels, depending on $m_a$.

%%%%%%%%%%%%%%%%%%%
\begin{figure}[t]
  \begin{center}
    \includegraphics[width=5.41cm]{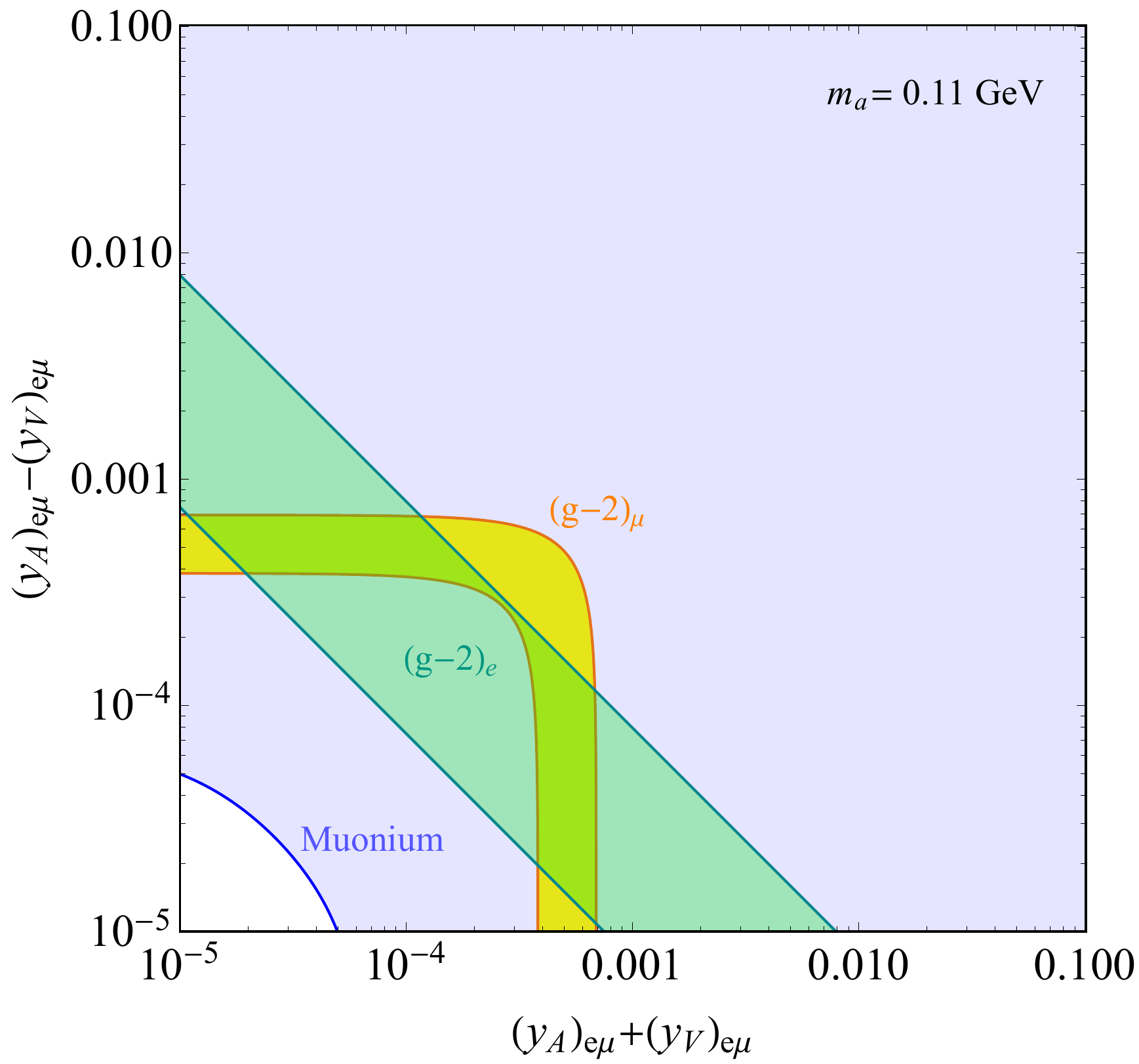}
      \includegraphics[width=5.41cm]{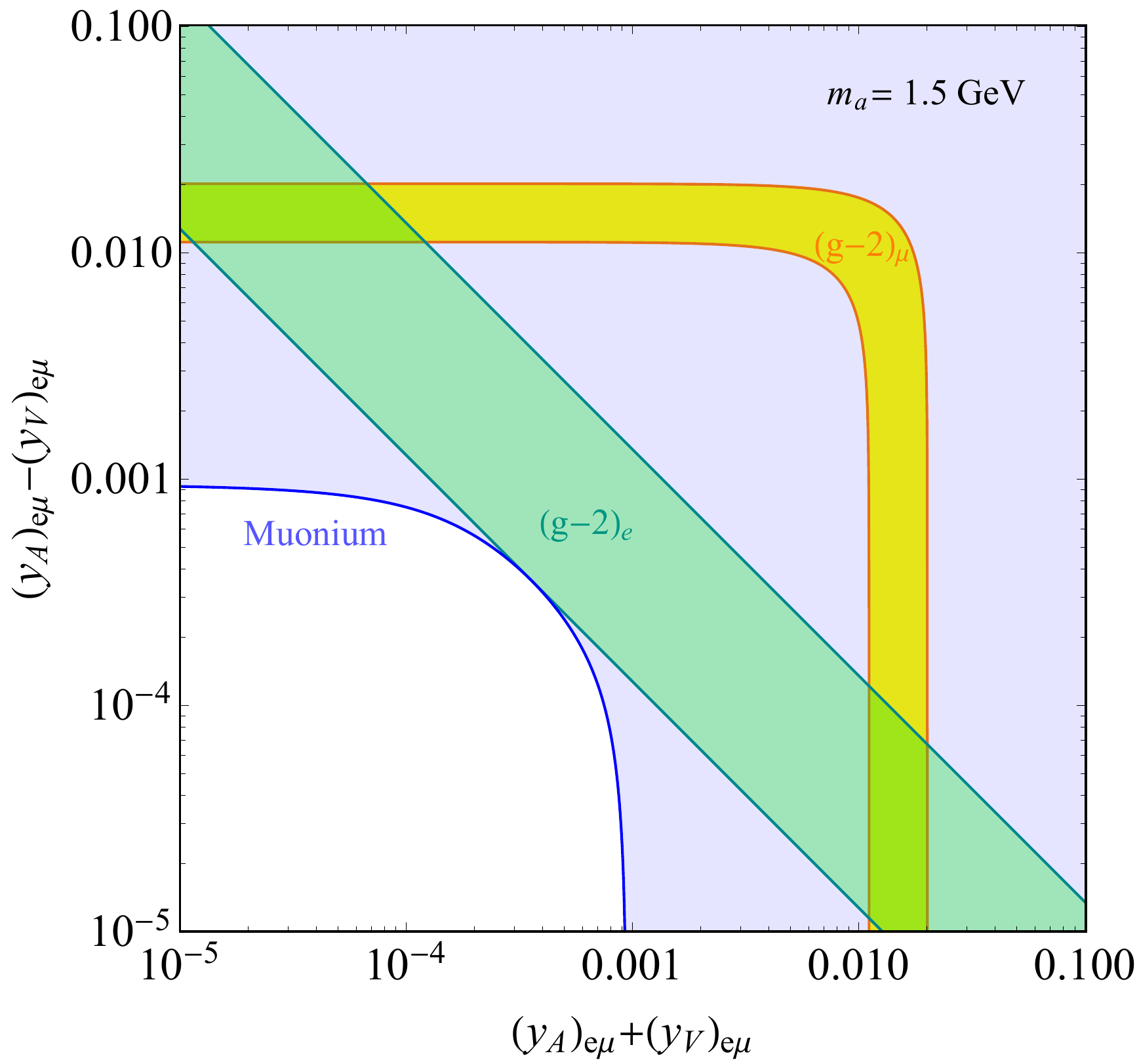}
      \includegraphics[width=5.41cm]{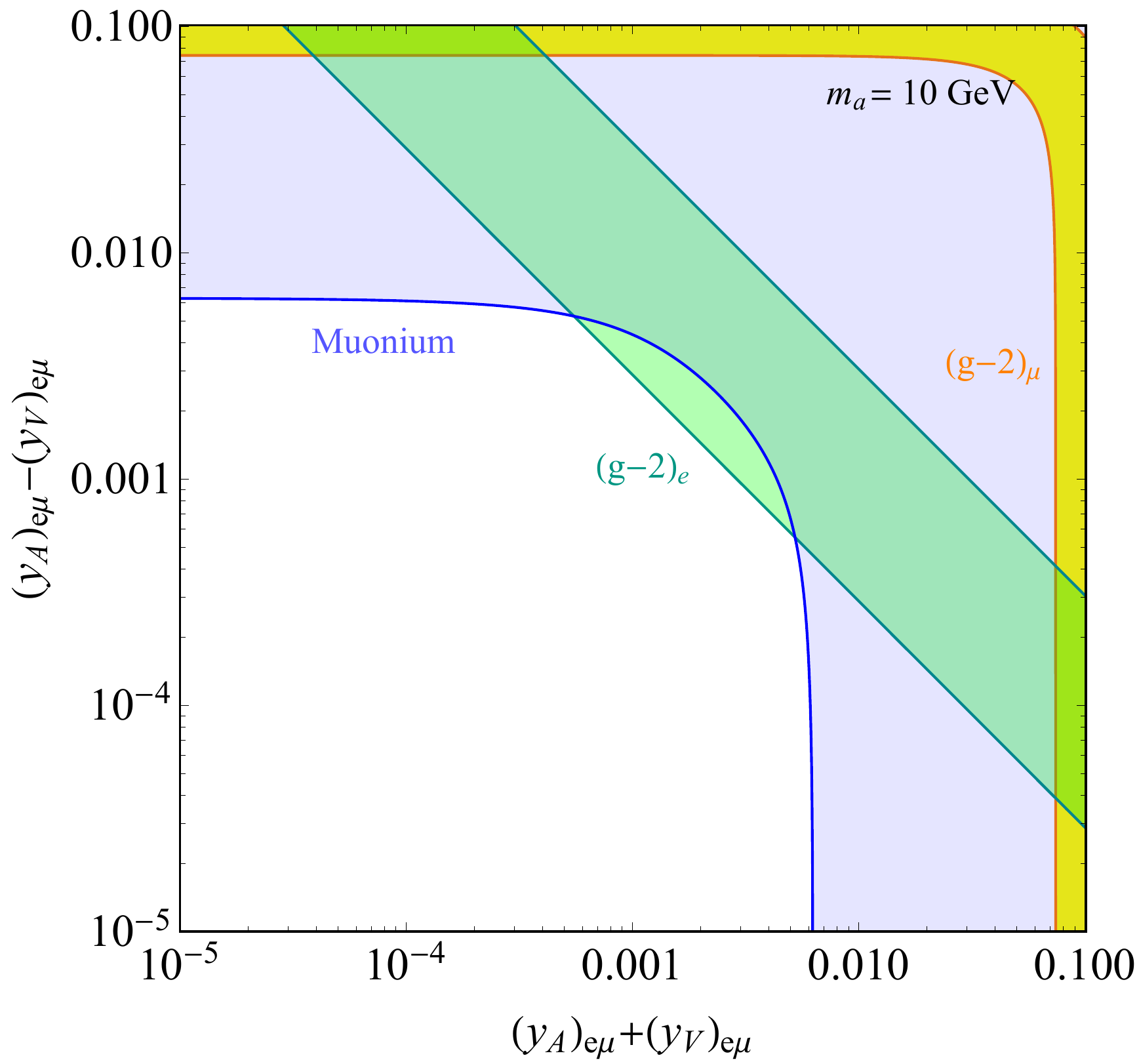}
            \caption{$\Delta a_e$ and $\Delta a_\mu$ are explained in the green and yellow regions within the $2\sigma$ level, respectively, 
            for $m_a = 110\,$MeV (left), $m_a = 1.5\,$GeV (middle) 
            and $m_a = 10\,$GeV (right).
            The blue-shaded regions  
            are excluded by the muonium-antimuonium oscillation bound at the $90\%$ CL.}
    \label{fig:muonium}
  \end{center}
\end{figure}
%%%%%%%%%%%%%%%%%%%%%

In Fig.~\ref{fig:muonium}, the $g-2$ favored regions are shown for $m_a = 110\,$MeV (left), $1.5\,$GeV (middle) and $10\,$GeV (right).  
The discrepancies of $(g-2)_{e}$ and $(g-2)_{\mu}$ are explained within the $2\sigma$ level in the green and yellow regions, respectively.
Here and hereafter, $\Delta a_\mu= \left( 27.8 \pm 7.4\right) \times 10^{-10}$ \cite{Keshavarzi:2019abf} is adopted. 
It is found that both anomalies can be reconciled by a mild tuning between $(y_V)_{\emu}$ and $(y_A)_{\emu}$ when the ALP is light, while the parameter tuning becomes tighter for larger $m_a$, as mentioned in Ref.~\cite{Bauer:2019gfk}.
In the figures, we also show the exclusion limit from the muonium-antimuonium oscillation measurement by the blue-shaded regions, which will be discussed in the next section.

When $m_a$ is smaller than $(m_\mu -m_e)$, the decay of $\mu \to e a$ is kinematically open and tightly constrained by a search for $\mu\to e+{\rm invisible}$ \cite{Bauer:2019gfk,Cornella:2019uxs,Bayes:2014lxz}.
Hence, no parameter space is available in $m_a < (m_\mu -m_e)$ to explain both of the anomalies.

Finally, let us comment on $CP$-violating contributions from the flavor-violating ALP couplings.
The most severe constraint comes from the electric dipole moment (EDM) of the electron, $|d_e| < 1.1 \times 10^{-29}\,e\,$cm ($90\%$ CL) \cite{Andreev:2018ayy}.
The contribution is generated by a similar diagram as Fig.~\ref{fig:diagram}, but the contribution is proportional to $\textrm{Im} \left[ (y_V)_{\emu} (y_A)_{\emu}^{\ast} \right]$.
Naively, it is estimated as $|d_e| \sim e /(2 m_e )a_e^{\rm ALP} 
\sim \mathcal{O}(10^{-23})\,e\,$cm, by requiring $a_e^{\rm ALP} \sim \Delta a_e$.
Therefore, the ALP couplings are required to satisfy $\textrm{Im} \left[ (y_V)_{\emu} (y_A)_{\emu}^{\ast} \right] < 10^{-6} \, [|\left(y_{A}\right)_{\emu}|^2- |\left(y_{V}\right)_{\emu}|^2]$.

%%%%%%%%%%%%%%%%%%%%%%%%%%%%%%%%%%%%
\section{Muonium-antimuonium oscillation}
\label{sec:muonium}
%%%%%%%%%%%%%%%%%%%%%%%%%%%%%%%%%%%%

Since the ALP is a real scalar boson, 
an exchange of the flavor-violating ALP generates effective $[\bar{\mu}e][\bar{\mu}e]$-type operators in a low-energy scale.
Such operators can be probed by measurements of a transition of a muonium (${\rm M}=\mu^+ e^-$) into an antimuonium ($\overline{\rm M}=\mu^- e^+$), which is so-called the muonium-antimuonium (${\rm M}\text{--}\overline{\rm M}$) oscillation \cite{Pontecorvo:1957cp,Feinberg:1961zz,Feinberg:1961zza}.
Since the SM prediction is highly suppressed by the neutrino masses \cite{Swartz:1989qz}, 
it is very sensitive to the $\emu$ flavor-violating interactions. 

According to Refs.~\cite{Hou:1995dg, Hou:1995np,Horikawa:1995ae}, 
we obtain the ${\rm M}\text{--}\overline{\rm M}$ transition probability
within the effective Lagrangian in Eq.~\eqref{eq:LeffYukawa} as
\beq
P_{{\rm M}\overline{\rm M}}= 
&\frac{8}{\pi^2 a_B^6 \lambda^2   m_a^4}\Biggl[
\left|c_{0,0}\right|^2 \left| \left(y_V\right)_{\emu}^2 - \left( 1 + \frac{1}{\sqrt{1+X^2}}\right) \left( y_A\right)^2_{\emu} \right|^2
\non 
&\quad \quad \quad \quad + \left|c_{1,0}\right|^2 \left| \left(y_V\right)_{\emu}^2 - \left( 1 - \frac{1}{\sqrt{1+X^2}}\right) \left( y_A\right)^2_{\emu} \right|^2\Biggr]\,,
\label{eq:muonium}
\eeq
where $\lambda = 3.00 \times 10^{-19}$\,GeV and $a_B = 2.69 \times  10^{5}$\,(GeV)$^{-1}$.
The factor $X$ in the right-handed side parametrizes effects of the external magnetic field $B$ as $X =6.31 B$\,(Tesla)$^{-1}$.  
The derivation is provided in Appendix~\ref{App:muonium}.
Note that this formula is obtained by integrating out the intermediate ALP, \ie, valid for $m_a > m_{\mu}$. 
We also assumed the $CP$ symmetry, and thus, parity-violating interferences between the scalar and pseudo-scalar interactions are dropped \cite{Glashow:1961zz}.

The most precise measurement has been performed by the MACS experiment
at PSI \cite{Willmann:1998gd}.
The upper bound on the ${\rm M}\text{--}\overline{\rm M}$ oscillation was obtained as
\beq
P_{{\rm M}\overline{\rm M}} < 8.3 \times  10^{-11}\,,
\label{eq:muoniumbound}
\eeq
at the $90\%$ CL.
In the experiment, the external magnetic field of $B= 0.1$\,Tesla 
was applied to detect an energetic $e^-$ from the antimuonium decay in a magnetic spectrometer. 
Besides, $c_{F,m_F} $ stands for a population of the muonium state in the experimental setup where $F$ is the total angular momentum of the muonium and $m_F= -F,-F+1,\dots, F$.
For the MACS experiment, it is estimated as $|c_{0,0}|^2 = 0.32$ and
$|c_{1,0}|^2 = 0.18$ \cite{Hou:1995dg,Hou:1995np}.\footnote{%
As a crosscheck, 
we reproduced magnetic field correction factors $S_B$ in Table II in Ref.~\cite{Willmann:1998gd} by supposing an equal population, $|c_{0,0}|^2=|c_{1,-1}|^2 = |c_{1,0}|^2= |c_{1,1}|^2 =0.25$. }
For specific cases, we obtain the upper bounds,\footnote{%
These bounds are consistent with those in Refs.~\cite{Harnik:2012pb,Dev:2017ftk}, where the magnetic effects are taken into account via $S_B$.
Also, our results are slightly severer than those in Refs.~\cite{Kim:1997rr,Evans:2019xer}.
}
\beq
\left|\left(y_V\right)_{\emu}\right| &< 
2.9
\times 10^{-4} \left(\frac{m_a}{\rm GeV} \right)\,, \quad  \textrm{for~}\left(y_A\right)_{\emu}=0\,,\\
\left|\left(y_A\right)_{\emu}\right| &<
2.4
\times 10^{-4}  \left(\frac{m_a}{\rm GeV} \right)\,, \quad  \textrm{for~}\left(y_V\right)_{\emu}=0\,,
\label{eq:muonium_axial}\\
\left|\left(y_V\right)_{\emu}\right|  &< 
3.1 \times 10^{-4} \left(\frac{m_a}{\rm GeV} \right) \,, \quad  \textrm{for~}\left(y_V\right)_{\emu}=\left(y_A\right)_{\emu}\,.
\eeq

In Fig.~\ref{fig:muonium}, the blue-shaded regions are excluded by the ${\rm M}\text{--}\overline{\rm M}$ oscillation.
It is shown that the region favored by the muon $g-2$ is completely excluded.\footnote{%
The same conclusion was made in Ref.~\cite{Dev:2017ftk} for a light scalar model with effective Yukawa couplings.}
Since both of the ALP contributions to $a_{\mu}$ and $P_{{\rm M}\overline{\rm M}}$ are scaled by $(y_{V,A})^2/m_a^2$ for $m_a \gg m_\mu$, $\Delta a_\mu$ cannot be explained even by a heavier ALP. 
On the other hand, $\Delta a_e$ can be explained for $m_a > 1.5$\,GeV.
Since the loop function of the ALP contribution to $a_e$ is proportional to $\log\left( m_a^2/m_\mu^2\right)/m_a^2$ for $m_a\gg m_\mu$, it is likely to be amplified compared to that for $a_\mu$, and the constraint from the ${\rm M}\text{--}\overline{\rm M}$ oscillation becomes relaxed. 
In the next section, we will study how to search for such an ALP at the Belle II experiment.

%%%%%%%%%%%%%%%%%%%%%%%%%%%%%%%%%%%%
\section{Collider signals at Belle II}
\label{sec:BelleII}
%%%%%%%%%%%%%%%%%%%%%%%%%%%%%%%%%%%%

In this section, we investigate experimental sensitivities
to the $\emu$ flavor-violating ALP at the Belle II experiment.
First, we consider a process $e^+ e^- \to \mu^{\pm} e^{\mp} a \to \mu^{\pm} \mu^{\pm} e^{\mp} e^{\mp}$, which is effective for $m_a \le \sqrt{s_{\rm{BelleII}}}$, where $\sqrt{s_{\rm{BelleII}}} = 10.58\,{\rm GeV}$ is the center-of-mass energy of the experiment.
Next, the FB asymmetry in the process $e^+ e^- \to \mu^+ \mu^-$ is investigated, where an off-shell ALP contributes, and thus, the observable has sensitivity to the ALP for $m_a \ge \sqrt {s_{\rm{BelleII}}}$.

%%%%%%%%%%%%%%%%%%%%%%%%%%%%%%%%%%%%
\subsection{
\texorpdfstring{\mbox{\boldmath$e^+ e^- \to  \mu^{\pm} \mu^{\pm}e^{\mp} e^{\mp}$}}{e e to mu mu e e} via on-shell ALP}
\label{sec:onshell}
%%%%%%%%%%%%%%%%%%%%%%%%%%%%%%%%%%%%

\begin{figure}[t]
  \begin{center}
 \includegraphics[width=5cm]{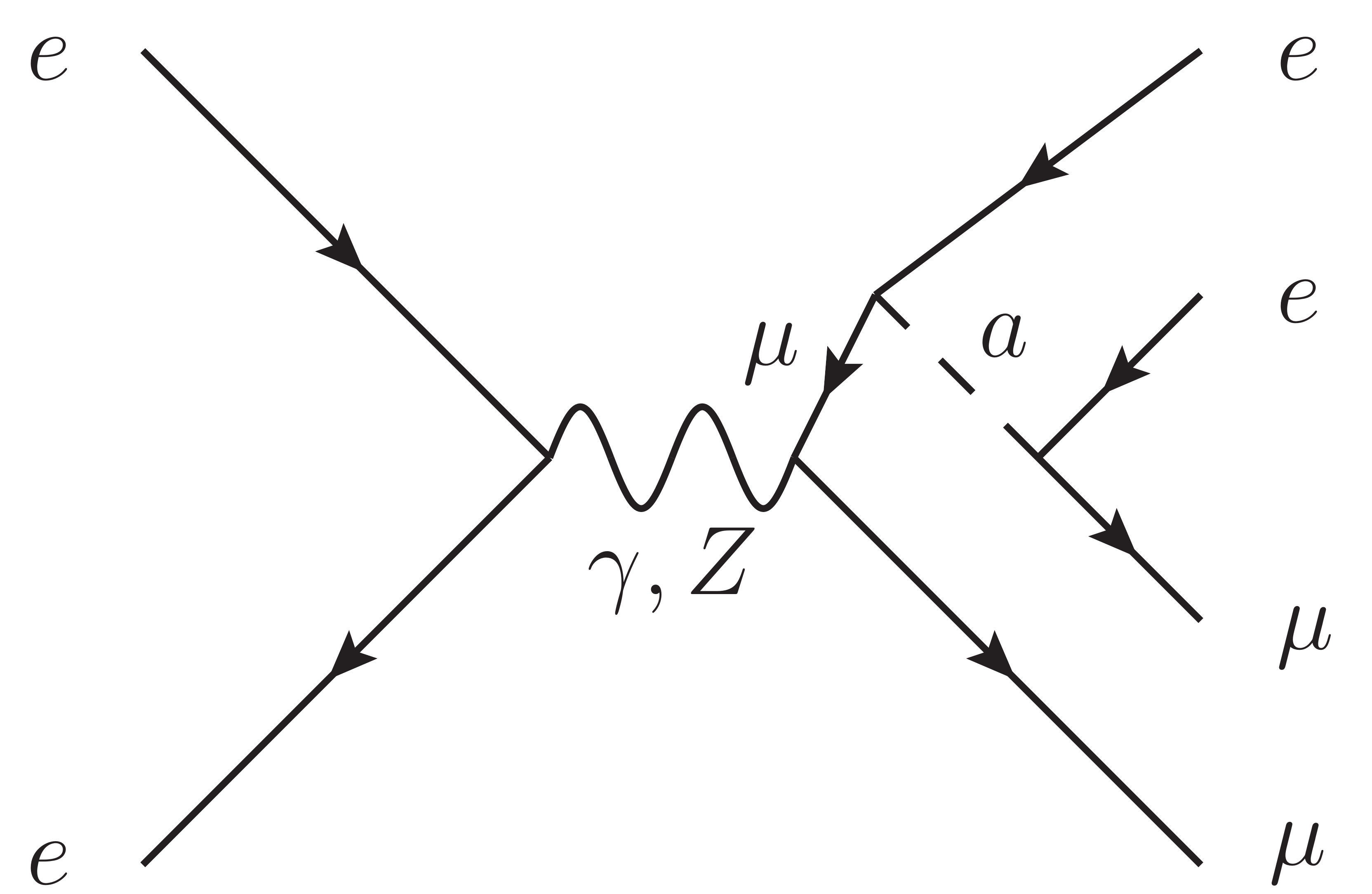}
 \quad
  \includegraphics[width=5cm]{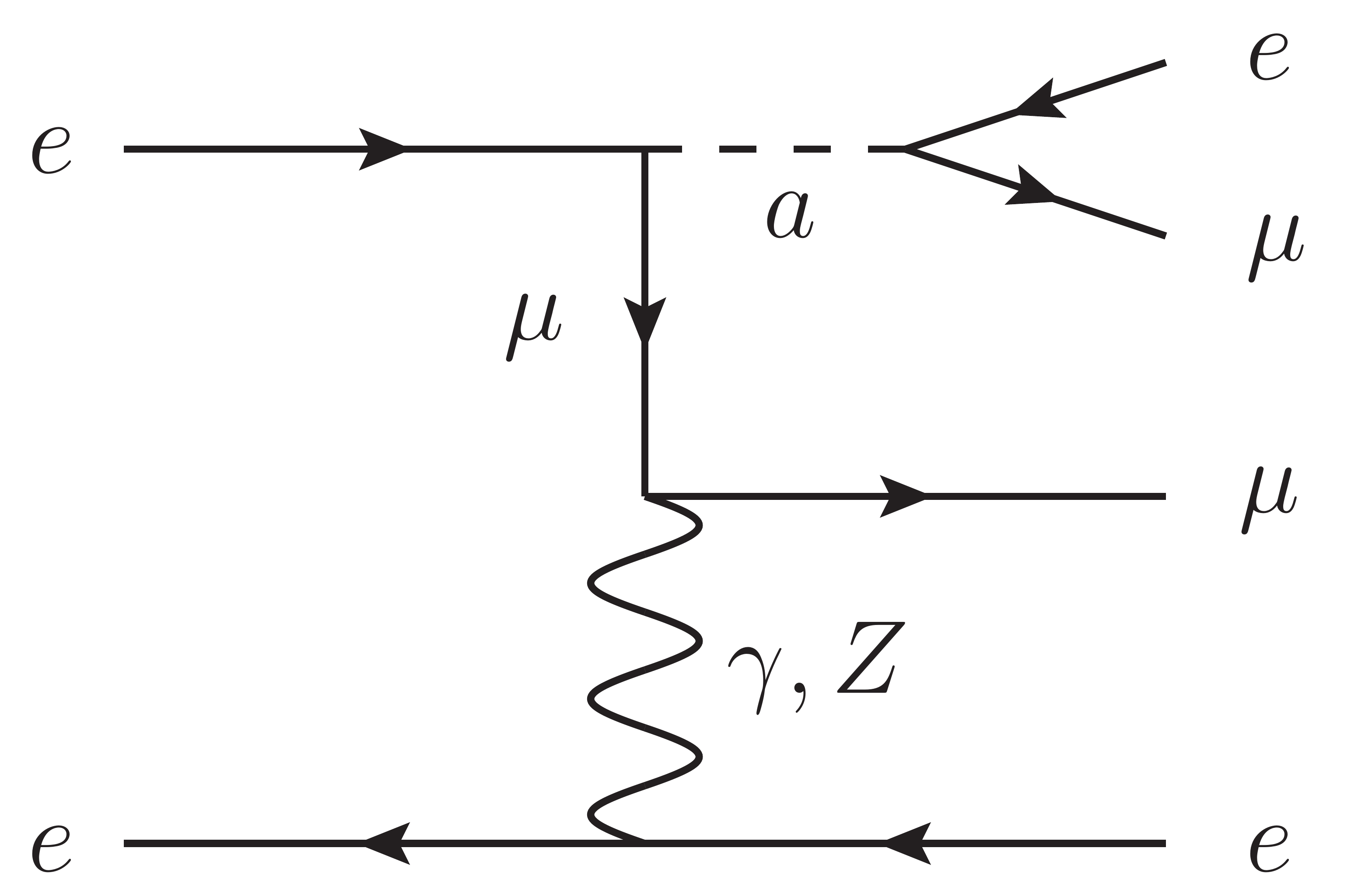}
 \quad 
  \includegraphics[width=5cm]{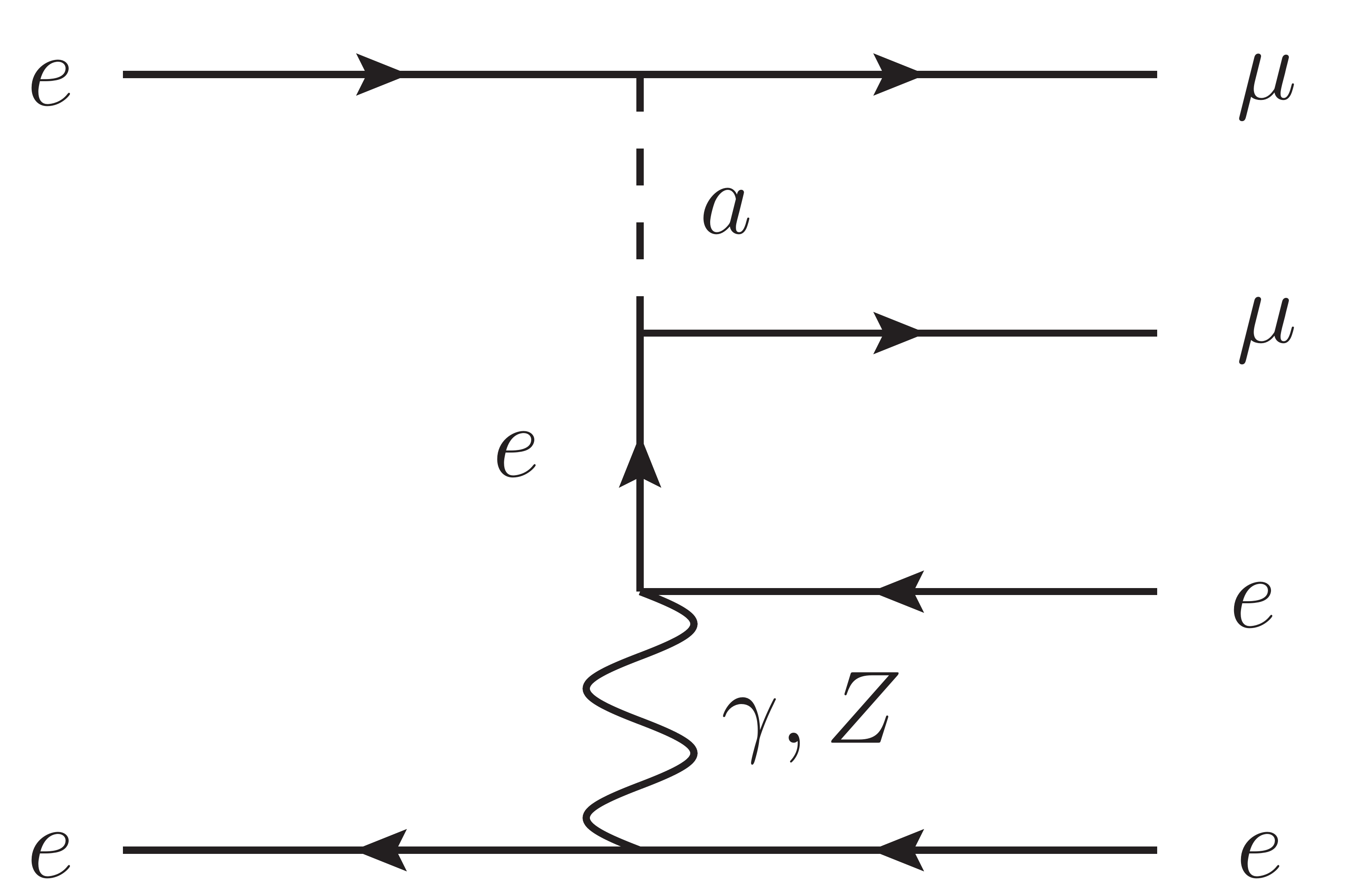}
            \caption{Examples of Feynman diagrams for $e^+ e^- 
\to \mu^{\pm} \mu^{\pm} e^{\mp} e^{\mp}$.
        }
    \label{diagramBelle}
  \end{center}
\end{figure}

\begin{figure}[t]
  \begin{center}
  \includegraphics[width=12cm]{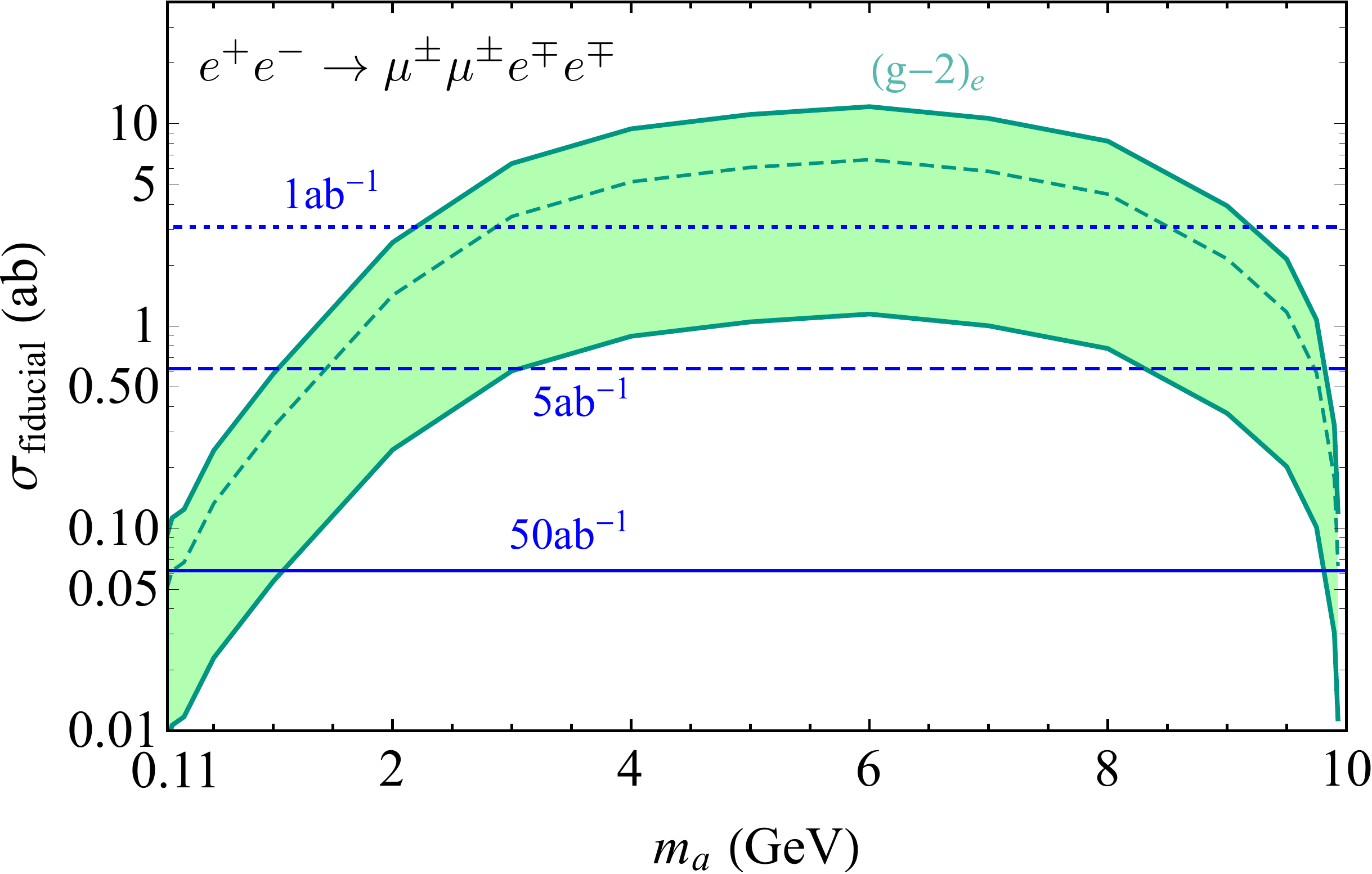}
        \caption{The fiducial production cross section of $e^+e^- \to \mu^{\pm} \mu^{\pm} e^{\mp}e^{\mp}$ as a function of $m_a$. Here, $(y_A)_{e\mu}$ and $m_a$ are varied, while $(y_V)_{e\mu}=0$ is fixed. 
        The electron $g-2$ discrepancy is explained within the $2\sigma$ level in the green band with the central value shown by the green dashed line. 
        The horizontal dotted, dashed, and solid blue lines represent the expected sensitivity of the Belle II experiment with the integrated luminosity of $1$, $5$, and $50$\,ab$^{-1}$, respectively.
        }
    \label{Xs}
  \end{center}
\end{figure}

In the $\emu$ flavor-violating ALP model, productions of the same-sign and same-flavor leptons pairs can proceed,\footnote{%
A process $e^+ e^- \to \mu^\pm e^\mp a \to \mu^\pm  \mu^\mp e^\pm e^\mp$ is also predicted but less distinctive due to larger backgrounds.} 
\beq
e^+ e^- \to \mu^\pm e^\mp a \to \mu^\pm  \mu^\pm e^\mp e^\mp \,.
\label{eq:process}
\eeq
The branching ratios of the ALP is $\textrm{BR}(a\to \mu^+ e^- ) = \textrm{BR}(a\to \mu^- e^+ )=0.5$ for $m_a \geq m_{\mu} + m_e$.
In Fig.~\ref{diagramBelle}, we show some of the diagrams which contribute to the process.
In the left and middle diagrams, an on-shell ALP is produced, while it is exchanged in an off-shell state in the right diagram.\footnote{%
In the analysis, although the off-shell contributions are included, we checked that the production cross section is dominated by the on-shell ALP productions.
Besides, ALP pair production, $ e^+ e^- \to a a \to \mu^\pm  \mu^\pm e^\mp e^\mp$, is included in our analysis, and we found that its contribution is negligible in Figs.~\ref{Xs} and \ref{fig:pure_pseudo}.
}

%%%%%%%%%%%%%%%%%%%
%%%%%%%%%%%%%%%%%%%%%

This process is quite unique;
the final state includes two pairs of the same-sign electrons and muons.
Such same-flavor and same-sign leptons processes are never produced within the SM.
Besides, the charge reconstruction of leptons is quite accurate in Belle II experiment \cite{Hanagaki:2001fz,Ikado:2006un,Kou:2018nap}. 
Hence, we simply neglect the SM background in the following analysis.\footnote{%
Another distinctive feature of the signal is an invariant mass peak around the ALP mass, $(p_e + p_{\mu})^2 =m_a^2$. 
Although there are two possible combinations to construct the $e\mu$ resonance, the wrong combination just provides a continuum distribution. To obtain a clear peak, we further need to drop it on event-by-event basis as discussed in Sec 3.3.1 of Ref.~\cite{Iguro:2019sly}.
In this paper, we do not impose this condition.
}

The fiducial production cross section is estimated by using {\tt MadGraph5} \cite{Alwall:2014hca}
with the ALP model file generated by {\tt FeynRules} \cite{Alloul:2013bka}.
Signal events with the asymmetric beam energy $E(e^-)=7$\,GeV and $E(e^+)=4$\,GeV \cite{Kou:2018nap}, are generated, 
and the following kinematical cuts for the final state leptons are imposed
in the laboratory frame \cite{Ikado:2006un,Kou:2018nap}:
\beq
e\,:& \quad 12^\circ \leq \theta \leq 155^\circ\,, \quad  \left|\vec{p}_e\right|\geq0.02\,\textrm{GeV}\,,\\
\mu\,:& \quad 25^\circ \leq \theta \leq 145^\circ\,,\quad  \left|\vec{p}_\mu\right|\geq0.6\,\textrm{GeV}\,,
\label{eq:cut}
\eeq 
where $\theta$ is an angle to the $e^-$ beam line in the laboratory frame.
Besides, the electron and muon tagging efficiencies are taken into account for each final-state lepton by following Ref.~\cite{Kou:2018nap}; they depend on the electron/muon energy but are assumed to be independent of the polar and azimuthal angles for simplicity.

%%%%%%%%%%%%%%%%%%%
%%%%%%%%%%%%%%%%%%%%%

Figure~\ref{Xs} shows the fiducial cross section after the kinematical cuts as a function of the ALP mass.
Here, $(y_A)_{e\mu}$ and $m_a$ are varied with $(y_V)_{e\mu}=0$ fixed.\footnote{%
When $(y_V)_{e\mu}\neq0$, larger couplings are required to explain $\Delta a_e$ 
because of the cancellation in Eq.~(\ref{eq:delta_ae}). 
Then, larger cross sections are predicted and the model can be tested with smaller luminosity.}
The green band represents a predicted cross section in which the electron $g-2$ discrepancy is explained within the $2\sigma$ level, and the green dashed line stands for the central value of the discrepancy.
The horizontal dotted, dashed, and solid  blue lines 
represent the expected sensitivities of the Belle II experiment with the integrated luminosity of $1$,\,$5$, and $50$\,ab$^{-1}$, respectively.
If no signal events are observed,
the region above the horizontal lines will be excluded at $95\%$ CL,
where the Poisson distribution is applied.
It is found that 
when the ALP accounts for the central value of $\Delta a_e$,
the Belle II experiment with the integrated luminosity of 5 (50)\,ab$^{-1}$
can test the model for $1.5 \leq m_a \leq 9.8$\,GeV ($0.15 \leq m_a \leq 9.9$\,GeV).
In a smaller mass region, the muon emitted by the ALP decay becomes soft and cannot pass the kinematical cut in Eq.~\eqref{eq:cut}.
On the other hand, 
the production cross section of $e^+ e^- \to \mu^\pm e^\mp a$ for a heavier ALP is suppressed both by the phase space and by the kinematical cuts.

Let us comment on a case with $m_a \gtrsim \sqrt {s_{\rm{BelleII}}}$.
Although the on-shell production of $a$ is kinematically forbidden,
the off-shell processes are still allowed.
However, 
since the production cross section becomes very small in the parameter region where the electron $g-2$ discrepancy is explained,
the ALP model will not be probed by the Belle II experiment via 
$e^+e^-
\to \mu^{\pm} \mu^{\pm} e^{\mp}e^{\mp}$.\footnote{%
Recently, LHC sensitivities to the flavor-violating ALP were studied by Ref.~\cite{Evans:2019xer} in a process of 
$pp\to h \to \mu^{\pm} \mu^{\pm} e^{\mp}e^{\mp} $.
It was shown that the region of $20 \lesssim m_a \lesssim 80$\,GeV can be probed accurately and its sensitivity is better than the ${\rm M}\text{--}\overline{\rm M}$ oscillation. 
Future lepton-collider sensitivities were also discussed in Ref.~\cite{Dev:2017ftk}.
}

%%%%%%%%%%%%%%%%%%%%%%%%%%%%%%%%%%%%
\subsection{Forward-backward asymmetry of 
\texorpdfstring{\mbox{\boldmath$e^+ e^- \to \mu^+ \mu^-$}}{e e to mu mu}}
\label{sec:AFB}
%%%%%%%%%%%%%%%%%%%%%%%%%%%%%%%%%%%%

Next, we consider the FB asymmetry of $e^+ e^- \to \mu^+ \mu^-$.
Within the SM, it occurs by an interference of $s$-channel $\gamma$ and $Z$ exchanges at the tree level.
Since the latter contribution is suppressed by a factor of $s_{\rm{BelleII}} /m_Z^2 \ll 1$, the SM value becomes tiny at the Belle II experiment \cite{Ferber:2015jzj,Ferber:2016rka}.
On the other hand, the flavor-violating ALP contributes in a $t$-channel process.
Such a contribution can modify an angular distribution of the muon pair and 
be probed by measuring the FB asymmetry:
\beq
A_{\rm FB} = 
\frac{N^+ \left( \cos \theta^{\ast} > 0 \right) -N^+ \left( \cos \theta^{\ast} < 0 \right)  }
{N^+ \left( \cos \theta^{\ast} > 0 \right) + N^+ \left( \cos \theta^{\ast} < 0 \right)  }\,,
\label{eq:AFBdef}
\eeq
where $\theta^{\ast}$ is an angle between $\mu^+$ and $e^+$ in the center-of-mass frame and $N^+\left( \cos \theta^{\ast} > 0 \right)$ is the number of $\mu^+$ satisfying $\cos \theta^{\ast} > 0$.

The ALP contribution to the FB asymmetry is evaluated by calculating an interference of the transition amplitudes of the SM and ALP contributions to the production cross section.
Since the SM amplitude is dominated by the $s$-channel $\gamma$ contribution, we obtain the interference term ${\cal I}$ as
\beq
{\cal I} & =  \frac{e^2}{2(k^2 - m_a^2) }
\left[ \left|\left(y_V\right)_{e \mu}\right|^2 + \left|\left(y_A\right)_{e \mu}\right|^2  \right]\non
&\quad \times  \left[  s  (1 + \cos^2 \theta^{\ast}) + 4 m_{\mu}^2(1 - \cos^2 \theta^{\ast})   - 4\sqrt{s\left( \frac{s}{4} - m_{\mu}^2\right)}\cos \theta^{\ast} \right]\,,
\eeq
where $s=s_{\rm{BelleII}}$ and $k$ is a momentum transfer of the ALP, 
\beq
k^2 &= m_{\mu}^2 - \frac{s}{2} + \sqrt{s\left( \frac{s}{4} - m_{\mu}^2 \right)} \cos \theta^{\ast}\,.
\label{eq:trans}
\eeq
The derivation is provided in Appendix~\ref{App:AFB}. 
Consequently, the ALP contribution to the FB asymmetry is given as
\beq
A_{\rm FB}^{\rm ALP} = 
 \frac{3}{8 e^{4} }
\left(
\int_{0}^1  d (\cos \theta^{\ast})\,  {\cal I}  - 
\int_{-1}^0 d (\cos \theta^{\ast})\,  {\cal I} \right)    \,.
\label{eq:AFBALPexact}
\eeq
When $| (y_{A})_{\emu}| \gg | (y_{V})_{\emu}|$
and $ m_a^2 \gg  s $,
it is approximated as
\beq
A_{\rm FB}^{\rm ALP} & \simeq 
  \frac{3}{32 \pi  \alpha}   
\left| \left(y_A\right)_{e \mu} \right|^2 \frac{ s }{m_a^2}\,.
\label{eq:AFBformula}
\eeq
If the Belle II experiment is supposed to detect $A_{\rm FB}^{\rm ALP} > \delta A_{\rm FB}$ in future, the ALP coupling 
\beq
\left| \left(y_A\right)_{e \mu} \right| > 
4.8 \times 10^{-2}
\sqrt{\delta A_{\rm FB}}
\left( \frac{m_a}{\rm GeV}\right)\,,
\label{eq:AFBBelle}
\eeq
could be probed
for $|(y_{A})_{\emu}| \gg | (y_{V})_{\emu}|$ and $m_a^2 \gg  s_{\rm{BelleII}}$.

%%%%%%%%%%%%%%%%%%%
\begin{figure}[t]
  \begin{center}
  \includegraphics[width=14cm]{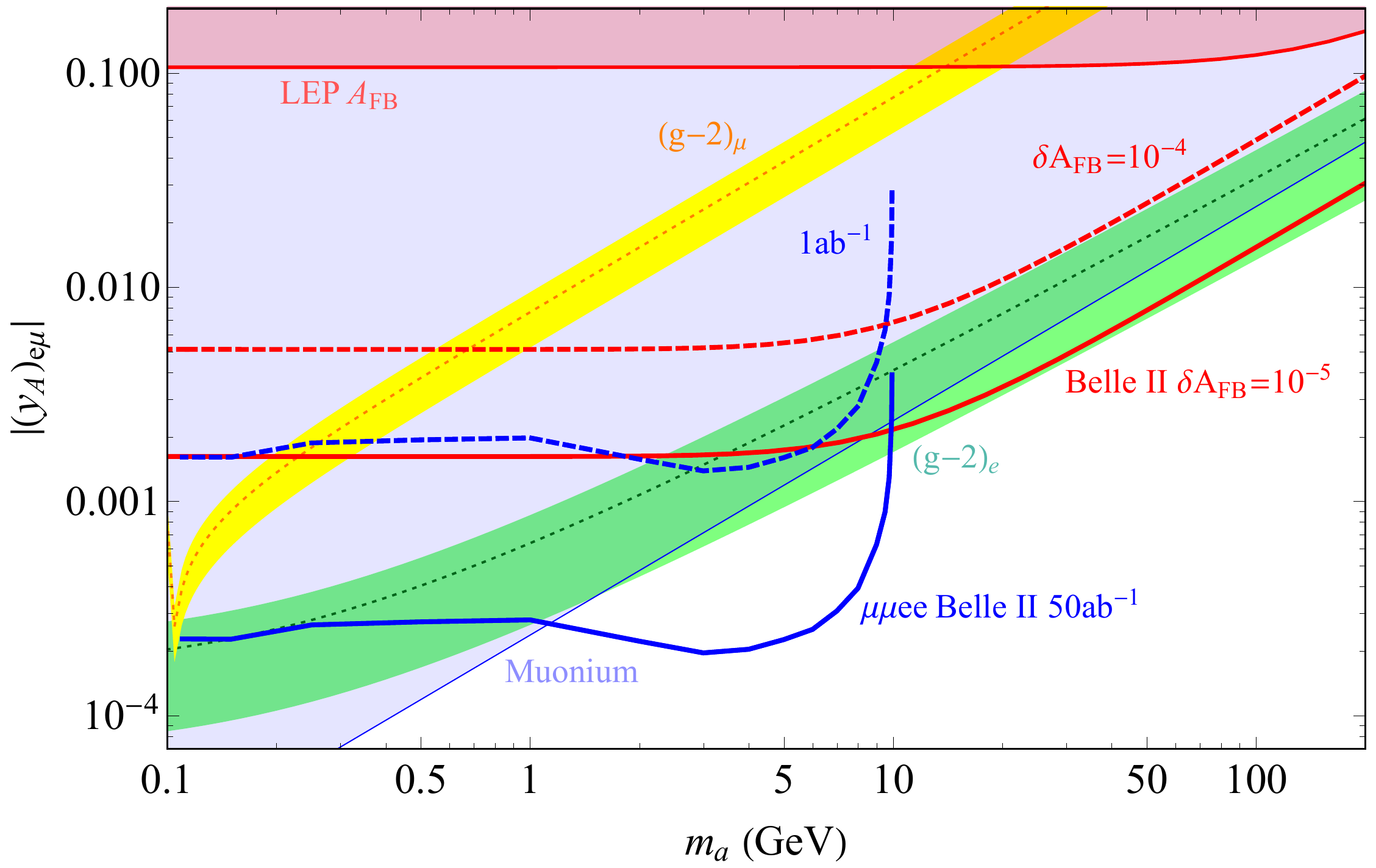}
            \caption{The Belle II sensitivities are shown with $(y_V)_{e \mu}=0$ fixed.
            The dashed and solid red lines represent the sensitivity of the $A_{\rm FB}$ measurement with resolutions of $\delta A_{\rm FB} = 10^{-4}$ and $10^{-5}$, respectively.
            The dashed and solid blue lines represent the sensitivity of searching for $e^+ e^- \to \mu^{\pm} \mu^{\pm}  e^{\mp}e^{\mp} $ with the integrated luminosity of $1$ and $50$\,ab$^{-1}$, respectively. 
            In the green and yellow regions, $\Delta a_e$ and $\Delta a_\mu$ are explained within the $2\sigma$ level with the central value drawn by the green and orange dotted lines, respectively.
            The blue-shaded region is excluded by the ${\rm M}\text{--}\overline{\rm M}$ oscillation at the $90\%$ CL.
            The red-shaded region is excluded by the LEP measurement of $A_{\rm FB}$.
            }
    \label{fig:pure_pseudo}
  \end{center}
\end{figure}
%%%%%%%%%%%%%%%%%%%%%

The Belle II experiment may achieve a statistical uncertainty of $\delta A_{\rm FB} \sim 10^{-5}$ with the integrated luminosity of 50\,ab$^{-1}$
\cite{Ferber:2015jzj}, though the systematic uncertainty is still unknown.
On the other hand, the uncertainty of the SM prediction is likely to be dominated by that of the vacuum polarization.\footnote{%
Although uncertainties from long-distance QED corrections could also be large, they might be suppressed by investigating $\mu^+ \mu^-$ angular distributions  
thanks to high statistics at the Belle II experiment \cite{Ferber:2016rka}.}
Its size is inferred to be 
as large as or slightly larger than the experimental statistical uncertainty by considering an analogy to a study of the FB asymmetry in the electroweak precision test (see, \eg, Ref.~\cite{deBlas:2016ojx} for a recent work).
Thus, reductions of the SM uncertainties could be essential to improve the Belle II sensitivity.
See Refs.~\cite{Jegerlehner:2006ju,Jegerlehner:2019lxt} for prospects of the vacuum polarization contribution.

In this paper, we consider $\delta A_{\rm FB} = 10^{-4}$ and $10^{-5}$ for the future sensitivity as a reference. %
In order to investigate the ALP contribution to the FB asymmetry at these accuracies, one needs ${\cal O}(10^{8-10})$ event samples for the Monte Carlo simulation. 
This is beyond the scope of this paper and we compare Eq.~\eqref{eq:AFBALPexact} with $\delta A_{\rm FB}$ to derive the future sensitivity.
Note that although signal acceptance and efficiencies should be taken into account, we have checked that such corrections are minor; the event number may be reduced by $\sim 10$\%.
We neglect them in the analysis for simplicity. 

In Fig.~\ref{fig:pure_pseudo}, 
the Belle II sensitivity to the FB asymmetry is shown for $\delta A_{\rm FB} = 10^{-4}$ and $10^{-5}$ by the dashed and solid red lines, respectively.
Here, $(y_V)_{\emu} =0$ is fixed.
In the green and yellow regions, $\Delta a_e$ and $\Delta a_\mu$ are explained within the $2\sigma$ level with the central value drawn by the green and orange dotted lines, respectively.
The ${\rm M}\text{--}\overline{\rm M}$ oscillation excludes the blue-shaded region at the $90\%$ CL.
It is seen that the region favored by the muon $g-2$ is completely excluded by the ${\rm M}\text{--}\overline{\rm M}$ oscillation. 
In contrast, although $\Delta a_e$ can be explained with satisfying the ${\rm M}\text{--}\overline{\rm M}$ oscillation bound, it is found that the Belle II measurement of the FB asymmetry with $\delta A_{\rm FB} = 10^{-5}$ can probe most of such parameter regions for $m_a \gtrsim 9\,$GeV.

The LEP experiment also performed measurements of the FB asymmetry of $e^+ e^- \to \mu^+ \mu^-$ for the center-of-mass energies from $130$\,GeV to $207$\,GeV \cite{Schael:2013ita}.
The experimental uncertainties are dominated by the statistical one, and one of the most precise results is provided at $\sqrt{s} = 207$\,GeV as 
\begin{align}
    A_{\rm FB}^{\rm LEP} = 0.535 \pm 0.028 \pm 0.004\,,
    \label{eq:LEPlimit}
\end{align}
where the first and second errors are statistical and systematic uncertainties, respectively.
On the other hand, it is referred to in Ref.~\cite{Schael:2013ita} that the SM value is 0.552 with the uncertainty smaller than the above statistical error.
Since this is consistent with the experimental result, the ALP contribution is constrained by the LEP experiment. 
In Fig.~\ref{fig:pure_pseudo},
we show the LEP $A_{\rm FB}$ bound at the $2\sigma$ level by the red-shaded region.\footnote{%
In the analysis, we include the $s$-channel $Z$ contribution to the SM amplitude as well as that from the $\gamma$ exchange, though the asymmetry is dominated by the interference of the ALP amplitude with the $\gamma$ contribution for $\sqrt{s} = 207$\,GeV.
}
It is found that the LEP measurement does not constrain the parameter region favored by the electron $g-2$.
Although the limit may be improved by combining all the LEP results in various center-of-mass energies, we expect that it is still weaker than the constraint by the ${\rm M}\text{--}\overline{\rm M}$ oscillation and cannot reach the Belle II sensitivities.\footnote{%
In Ref.~\cite{Schael:2013ita}, constraints for four-fermion interactions are also investigated by combining all the LEP results; 
according to Table 3.15, one finds $\Lambda > 12.1\,$TeV for A$0^{-}$ model, which corresponds to a purely pseudo-scalar electron-muon interaction.
See the reference for definitions of the parameters. 
Since $\left( y_A\right)_{\emu}$ is related to $\Lambda$ as $| ( y_A )_{\emu} | = \sqrt{8 \pi} m_a/\Lambda$, the limit is converted to be
\beq
  \left|\left( y_A\right)_{\emu}\right| 
 < 0.41 \left(\frac{m_a}{\rm TeV}\right)\,,
 \label{eq:LEP}
\eeq
at the $95\%$ CL, where the ALP is assumed to be decoupled. 
This result is consistent with  Ref.~\cite{Dev:2017ftk}.
On the other hand, by using Eq.~\eqref{eq:LEPlimit} we obtain
\beq
  \left|\left( y_A\right)_{\emu}\right| 
 < 0.57 \left(\frac{m_a}{\rm TeV}\right)\,,
\eeq
for $m_a \gg 207\,$GeV.
Thus, our result relying solely on Eq.~\eqref{eq:LEPlimit} is not much weaker than the fully combined limit, Eq.~\eqref{eq:LEP}.
} 

Finally, in Fig.~\ref{fig:pure_pseudo}, 
the Belle II sensitivity to a search for $e^+ e^- \to \mu^\pm e^\mp a \to \mu^\pm  \mu^\pm e^\mp e^\mp$, which was discussed in the last subsection, is also shown for the integrated luminosity of $1$ and $50$\,ab$^{-1}$ by the dashed and solid blue lines, respectively.
For $1\lesssim m_a \lesssim 10$\,GeV with $50$\,ab$^{-1}$, it is found that the process is useful to probe the parameter region which is favored by the electron $g-2$ and consistent with the ${\rm M}\text{--}\overline{\rm M}$ oscillation data. 
This result is complementary to the FB asymmetry; we conclude that the ALP parameter region favored by the electron $g-2$ could be tested almost entirely by the Belle II experiment.

%%%%%%%%%%%%%%%%%%%%%%%%%%%%%%%%%%%%
\section{Conclusion}
\label{sec:conclusion}
%%%%%%%%%%%%%%%%%%%%%%%%%%%%%%%%%%%%

In this paper,
we revisited the $\emu$ flavor-violating ALP model motivated by the electron and muon $g-2$ anomalies.
Such an ALP inevitably induces the muonium-antimuonium oscillation, and it was found that whole the parameter regions which explain both of the discrepancies simultaneously are already excluded.

Nevertheless, the model can accommodate only the  electron $g-2$ anomaly.
In order to test such parameter regions, we investigated the Belle II sensitivity of searching for the on-shell ALP production of $e^+ e^- \to \mu^\pm e^\mp a$ followed by $a \to \mu^\pm e^\mp$ and the forward-backward asymmetry of $e^+ e^- \to \mu^+ \mu^-$.
We found that the former provides a good sensitivity for $m_a < 10\,$GeV, 
and the latter does for $m_a > 10\,$GeV.
Hence, these measurements are complementary and provide better sensitivities than the muonium-antimuonium oscillation bound for $m_a > 1\,$GeV.
As a result, the parameter region favored by the electron $g-2$ anomaly can be tested almost entirely by the Belle II experiment with the integrated luminosity of $50$\,ab$^{-1}$.

Finally, 
we briefly comment on a light and flavor-violating complex scalar field.
If this complex scalar $S$ carries an electron or muon charge, 
the low-energy effective interactions become
$- (\partial_{\mu} S/\Lambda ) \bar{e}\gamma^{\mu} (v_{e\mu} -a_{e\mu}\gamma_5 )\mu$ 
or $(y_V)_{e\mu}S  \bar{e}\mu +  (y_A)_{e\mu}S  \bar{e}\gamma_5\mu$ with their Hermitian conjugate, while interactions of $e \leftrightarrow \mu$ are forbidden.
Although the contributions to the lepton $g-2$ become analogous to Eqs.~\eqref{eq:delta_ae} and \eqref{eq:delta_amu}, the transition of the muonium into antimuonium does not proceed via the scalar.
Hence, the flavor-violating complex scalar can explain both $g-2$ anomalies simultaneously, though the UV completion is nontrivial \cite{Evans:2019xer}.\footnote{%
Note that when the effective interaction is $(y_R)_{e\mu}S  \bar{e}P_R \mu  + \textrm{H.c.}$, the contribution to the electron $g-2$ vanishes, while the muon $g-2$ anomaly can be explained.}\footnote{%
Even when there is a mass difference between the scalar and pseudo-scalar components of the complex scalar field, 
a destructive interference works between their contributions 
to the muonium-antimuonium oscillation.
In particular, 
when a phase component in the complex scalar corresponds to ALP,
its contribution to the 
oscillation could be diminished by a scalar partner 
unless its mass is far from the ALP one.}
At the Belle II experiment, the measurement of the forward-backward asymmetry could probe such a parameter region when $m_S > \mathcal{O}(1)$\,GeV;
the muon $g-2$ discrepancy requires larger couplings, which enhance $A_{\rm{FB}}$, too.
Although the same-flavor and same-sign productions of the electron and muon pairs do never proceed, one can consider a flavor-violating resonant search in the same-flavor and opposite-sign lepton pair productions of $e^+ e^- \to \mu^\pm e^\mp a \to \mu^\pm e^\mp \mu^\mp e^\pm$, which could be a target in the early stage of the Belle II experiment.

%%%%%%%%%%%%%%%%%%%%%%%%%%%%%%%%%%%%
%%%%%%%%%%%%%%%%%%%%%%%%%%%%%%%%%%%%
\section*{Acknowledgements}
We would like to thank 
Yael Shadmi and Yotam Soreq 
for worthwhile discussions of the forward-backward asymmetry 
in the muon pair production.
We also thank 
Iftah Galon, Junji Hisano, Kenji Inami, Akimasa Ishikawa, Kodai Matsuoka, Satoshi Mishima, Tsuzuki Noritsugu, Olcyr Sumensari, Kazuhiro Tobe, and Koji Tsumura 
for many useful comments and discussions.
S.I.\ would like to thank Michigan State University for its warm hospitality where this project was initiated. 
%---------------------------------------------------------------------------
The work of S.I.\ is supported by Kobayashi-Maskawa Institute for the Origin of Particles and the Universe, Toyoaki scholarship foundation 
and the Japan Society for the Promotion of Science (JSPS) Research Fellowships for Young Scientists, No. 19J10980.
%%%
%%%
This work is supported in part by the Grant-in-Aid for 
Scientific Research B (No.16H03991 [ME]), and 
Early-Career Scientists (No.16K17681 [ME] and No.19K14706 [TK]).
The work of T.K.\ is also supported by 
the Israel Science Foundation (Grant No.~751/19).
%---------------------------------------------------------------------------

%%%%%%%%%%%%%%%%%%%%%%%%%%%%%%%%%%%%
%%%%%%%%%%%%%%%%%%%%%%%%%%%%%%%%%%%%

\appendix

\section{Muonium-antimuonium transition}
\label{App:muonium}

In this appendix, 
a probability of the transition of the muonium (M $= \mu^+ e^-$) into antimuonium ($\overline{\rm M} =\mu^- e^+$) is derived under an external magnetic field $B$.
We follow the analysis explored in Ref.~\cite{Hou:1995np}.

Neglecting effects of the $B$ field, depending on spins of the leptons, there are four types of the 
$1S$ muonium state $|$M; $J,m_J\rangle$:
\beq
& |{\rm M};0,0\rangle_0 = \frac{1}{\sqrt{2}}\left( - |{\rm M};\uparrow,\downarrow\rangle + |{\rm M};\downarrow,\uparrow\rangle 
\right)\,,\\
& |{\rm M};1,0\rangle_0 = \frac{1}{\sqrt{2}}\left(  |{\rm M};\uparrow,\downarrow\rangle + |{\rm M};\downarrow,\uparrow\rangle 
\right)\,,\\
& |{\rm M};1,1\rangle =   |{\rm M};\uparrow,\uparrow\rangle\,,
\quad \quad |{\rm M};1,-1\rangle= 
|{\rm M};\downarrow,\downarrow\rangle\,.
\eeq
Under the nonzero external $B$ filed, they are modified via the magnetic dipole moment of leptons:
\beq
|{\rm M};0,0\rangle_B&=\frac{ c+s}{\sqrt{2}}  |{\rm M};0,0\rangle_0 
+ \frac{ c-s}{\sqrt{2}} |{\rm M};1,0\rangle_0\,, 
\label{eq:B1}
\\
|{\rm M};1,0\rangle_B&=\frac{ c+s}{\sqrt{2}}  |{\rm M};1,0\rangle_0 
- \frac{ c-s}{\sqrt{2}} |{\rm M};0,0\rangle_0\,,
\label{eq:B2}
\eeq
and the $|{\rm M};1,\pm1\rangle$ states change only their energy levels.
Here, $s$ and $c$ are
\begin{align}
 s = \frac{1}{\sqrt{2}} \left[ 1 - \frac{X}{\sqrt{1+X^2}} \right]^{\frac{1}{2}}\,,
 \quad
 c = \frac{1}{\sqrt{2}} \left[ 1 + \frac{X}{\sqrt{1+X^2}} \right]^{\frac{1}{2}}\,,
\end{align}
with a dimensionless parameter,
\begin{align}
 X = \frac{\mu_B B}{a} \left( g_e + \frac{m_e}{m_\mu} g_\mu \right) \simeq 6.31 \,  \frac{B}{{\rm Tesla}}\,,
\end{align}
where $\mu_B = e/(2 m_e)$ is the Bohr magneton, $g_e \simeq g_{\mu}\simeq 2$ is the magnetic moment of the electron/muon, and $a \simeq 1.846 \times 10^{-5}\,$eV is the $1S$ muonium hyperfine splitting.
Similarly, the antimuonium sates are represented as
\beq
|\overline{\rm M};0,0\rangle_B&=\frac{ c+s}{\sqrt{2}}  |\overline{\rm M};0,0\rangle_0 
- \frac{ c-s}{\sqrt{2}} |\overline{\rm M};1,0\rangle_0\,, 
\label{eq:B3}
\\
|\overline{\rm M};1,0\rangle_B&=\frac{ c+s}{\sqrt{2}}  |\overline{\rm M};1,0\rangle_0 
+ \frac{ c-s}{\sqrt{2}} |\overline{\rm M};0,0\rangle_0\,. 
\label{eq:B4}
\eeq

The transition probability of M $\to \overline{\rm M}$ under the nonzero $B$ field is denoted as
\beq
P_{{\rm M}\overline{\rm M}}\simeq |c_{0,0}|^2 P^{(0,0)}_{{\rm M}\overline{\rm M}} 
+ |c_{1,0}|^2 P^{(1,0)}_{{\rm M}\overline{\rm M}}\,,
\label{eq:muoniumprobability}
\eeq
where $|c_{J,m_J}|$ is the population probability of the muonium initial state with $(J,m_J)$, which satisfies $|c_{0,0}|^2+|c_{1,-1}|^2+|c_{1,0}|^2+|c_{1,1}|^2 =1$.
Note that the transitions of $|{\rm M};0,0\rangle_B \to |\overline{\rm M};1,0\rangle_B$ and $|{\rm M};1,0\rangle_B \to |\overline{\rm M};0,0\rangle_B$
are extremely suppressed due to the hyperfine splitting, and that of 
$|{\rm M};1,\pm1 \rangle_B \to |\overline{\rm M};1,\pm1 \rangle_B$ is also suppressed because the energy levels of the initial and final states are modified under the nonzero magnetic field.

The transition probabilities are represented by the transition amplitudes as \cite{Feinberg:1961zz,Feinberg:1961zza},
\beq
P^{(0,0)}_{{\rm M}\overline{\rm M}}  \simeq \frac{2}{\lambda^2}
\left| \langle \overline{\rm M};0,0 | \mathcal{H}_{\rm eff} | {\rm M}; 0,0 \rangle_B \right|^2\,, 
\quad 
P^{(1,0)}_{{\rm M}\overline{\rm M}}  \simeq \frac{2}{\lambda^2}
\left| \langle \overline{\rm M};1,0 | \mathcal{H}_{\rm eff} | {\rm M}; 1,0 \rangle_B \right|^2\,,
\eeq
where $\lambda = (\tau_{\mu})^{-1} \simeq 3.00\times10^{-19}$\,GeV is the muon decay rate.
Using Eqs.~\eqref{eq:B1}--\eqref{eq:B4}, 
the transition amplitudes are written as
\beq
 \langle \overline{\rm M};0,0 | \mathcal{H}_{\rm eff} | {\rm M}; 0,0 \rangle_B
 = \frac{1+ 2 s c }{2}\langle \overline{\rm M};0,0 | \mathcal{H}_{\rm eff} | {\rm M}; 0,0 \rangle_0
 - \frac{1- 2 s c }{2}\langle \overline{\rm M};1,0 | \mathcal{H}_{\rm eff} | {\rm M}; 1,0 \rangle_0\,,\\
  \langle \overline{\rm M};1,0 | \mathcal{H}_{\rm eff} | {\rm M}; 1,0 \rangle_B
 = - \frac{1- 2 s c }{2}\langle \overline{\rm M};0,0 | \mathcal{H}_{\rm eff} | {\rm M}; 0,0 \rangle_0
 + \frac{1+ 2 s c }{2}\langle \overline{\rm M};1,0 | \mathcal{H}_{\rm eff} | {\rm M}; 1,0 \rangle_0\,.
\eeq

In the case of the effective Lagrangian of Eq.~\eqref{eq:LeffYukawa},
when the ALP is sufficiently heavier than the muonium, 
the effective Hamiltonian is obtained by integrating the ALP field as
\begin{align}
 \mathcal{H}_{\rm eff} =  - \frac{\left(y_V\right)_{\emu}^2}{m_a^2} S^2 -  \frac{\left( y_A\right)_{\emu}^2}{m_a^2} P^2\,,
\end{align}
where $S = \bar\mu e$ and $P = \bar\mu \gamma_5 e$. 
Here, the $CP$ conservation is assumed.
Neglecting effects of the magnetic field, the transition matrix elements are
\cite{Hou:1995dg}
\begin{align}
 &
 \langle \overline{\rm M}; F=0 | S^2 | {\rm M}; F=0 \rangle_0 = \frac{2}{\pi a_B^3}\,,~~~
 \langle \overline{\rm M}; F=1 | S^2 | {\rm M}; F=1 \rangle_0 = - \frac{2}{\pi a_B^3}\,,
 \\ &
 \langle \overline{\rm M}; F=0 | P^2 |  {\rm M}; F=0 \rangle_0 = - \frac{4}{\pi a_B^3}\,,~~~
 \langle \overline{\rm M}; F=1 | P^2 | {\rm M}; F=1 \rangle_0 = 0\,,
\end{align}
where $a_B = 1/(m_r \alpha)\simeq 2.69\times 10^{5}$\,(GeV)$^{-1}$ is the Bohr radius,
and $m_r = m_e m_{\mu}/(m_e + m_{\mu})$ is the reduced mass.
Thus, we obtain
\begin{align}
\langle \overline{\rm M};0,0 | \mathcal{H}_{\rm eff} | {\rm M}; 0,0 \rangle_0& 
= 
 - \frac{\left(y_V\right)_{\emu}^2}{m_a^2} \frac{2}{\pi a_B^3} + \frac{\left(y_A\right)_{\emu}^2}{m_a^2} \frac{4}{\pi a_B^3}
\, ,
 \\
\langle \overline{\rm M};1,0 | \mathcal{H}_{\rm eff} | {\rm M}; 1,0 \rangle_0& 
= 
  \frac{\left(y_V\right)_{\emu}^2}{m_a^2} \frac{2}{\pi a_B^3}\,.
\end{align}

Consequently, the transition probabilities is derived as
\beq
P^{(0,0)}_{{\rm M}\overline{\rm M}}
& = 
 \frac{8}{\pi^2 a_B^6 \lambda^2} 
 \left| \frac{\left(y_V\right)_{\emu}^2}{m_a^2} - \left(1 +  \frac{1}{\sqrt{1+X^2}}\right) \frac{\left(y_A\right)_{\emu}^2}{m_a^2} \right|^2\,,\\
 P^{(1,0)}_{{\rm M}\overline{\rm M}}
& = 
 \frac{8}{\pi^2 a_B^6 \lambda^2} 
 \left| \frac{\left(y_V\right)_{\emu}^2}{m_a^2} - \left(1 -  \frac{1}{\sqrt{1+X^2}}\right) \frac{\left(y_A\right)_{\emu}^2}{m_a^2} \right|^2\,.
\eeq
Substituting them into Eq.~\eqref{eq:muoniumprobability},
one obtains Eq.~\eqref{eq:muonium}.

\section{Forward-backward asymmetry}
\label{App:AFB}

In this appendix,
the ALP contribution to the FB asymmetry of $e^+ e^- \to \mu^+ \mu^-$ is calculated.
The asymmetry is obtained by an interference of the SM and ALP scattering amplitudes. 
In the former, a contribution with the virtual $\gamma$ exchange dominates the amplitude for the energy $s \ll m_Z^2$.
Then, it becomes
\beq
\mathcal{M}_{\gamma} = \frac{ e^2}{s} (\bar v_{e^+} \gamma^{\mu} u_{e^-})  (\bar u_{\mu^-} \gamma_{\mu} v_{\mu^+})\,.
\eeq
We define 
$\theta^{\ast}$ as the angle between $\mu^+$ and $e^+$ in the center-of-mass frame.
The integration of the squared amplitude over $\cos \theta^{\ast}$ including a factor of the polarization sums leads to
\beq
\int_{-1}^1 d(\cos \theta^{\ast})\, \left(\frac{1}{2}\right)^2 |\mathcal{M}_{\gamma}|^2 
= \frac{8}{3} e^4 \left(1 + \frac{2m_{\mu}^2}{  s}  \right)\,,
\label{eq:AFBdenom}
\eeq
which gives a normalization for $A_{\rm FB}$.
Note that $|\mathcal{M}_{\gamma}|^2$ does not generate $A_{\rm FB}$.
On the other hand, 
the ALP amplitude is obtained as
\beq
\mathcal{M}_{a}
& =  \frac{1}{2(k^2 - m_a^2)}
\biggl\{\left|\left(y_V\right)_{e \mu} - \left(y_A\right)_{e \mu} \right|^2 
(\bar v_{e^+} \gamma^{\mu} P_R u_{e^-}) (\bar u_{\mu^-} \gamma_{\mu} P_L v_{\mu^+})
\non
& \quad \quad  \quad  \quad  \quad \quad  +\left|\left(y_V\right)_{e \mu} +  \left(y_A\right)_{e \mu}\right|^2 
(\bar v_{e^+}  \gamma^{\mu} P_L u_{e^-} ) (\bar u_{\mu^-} \gamma_{\mu}  P_R v_{\mu^+})\biggr\}
\non
&\quad  +    \frac{1}{2(k^2 - m_a^2)}
\biggl\{ \left[ \left|\left(y_V\right)_{e \mu}\right|^2 - \left|\left(y_A\right)_{e \mu}\right|^2 + 2 i {\rm Im}  \left(y_V\right)_{e \mu} \left(y_V\right)_{e \mu}^{\ast}\right]\non
&\quad \quad \quad \quad \times
\biggl[ 
(\bar v_{e^+}  P_L u_{e^-}) (\bar u_{\mu^-} P_L v_{\mu^+} )  + \frac{1}{4} (\bar v_{e^+}  \sigma^{\mu \nu}  P_L u_{e^-}) (\bar u_{\mu^-}  \sigma_{\mu \nu}P_L v_{\mu^+})\biggr]\non
& \quad +  \left[  \left|\left(y_V\right)_{e \mu}\right|^2 - \left|\left(y_A\right)_{e \mu}\right|^2 -  2 i {\rm Im}  \left(y_V\right)_{e \mu} \left(y_V\right)_{e \mu}^{\ast}\right]\non
&\quad \quad  \quad \quad \times
\biggl[ (\bar v_{e^+}  P_R u_{e^-}) (\bar u_{\mu^-} P_R v_{\mu^+} ) 
+ \frac{1}{4}  (\bar v_{e^+} \sigma^{\mu \nu} P_R u_{e^-}) (\bar u_{\mu^-}   \sigma_{\mu \nu}P_R v_{\mu^+} )\biggr] \biggr\}\,,
\label{eq:amplitudea}
\eeq
after the Fierz rearrangement of the fermion order.
Here, $ \sigma^{\mu\nu} = \frac{i}{2} [\gamma^{\mu}, \gamma^{\nu}]$
and
\beq
k^2 &= m_{\mu}^2 - \frac{s}{2} + 
\sqrt{s\left( \frac{s}{4} - m_{\mu}^2 \right)} \cos \theta^{\ast}\,.
\label{eq:ksq}
\eeq
Discarding the electron mass, 
we obtain the interference term of $\gamma$ and $a$ as
\beq
{\cal I} & = \left(\frac{1}{2}\right)^2 
2 \textrm{Re} \left(  \mathcal{M}_{\gamma} \mathcal{M}_{a}^{\ast}\right)
\non\\
%%%
& =  \frac{e^2}{2(k^2 - m_a^2) }
\left[ \left|\left(y_V\right)_{e \mu} \right|^2 + \left|\left(y_A\right)_{e \mu}\right|^2   \right]\non
&\quad \times  \left[  s  (1 + \cos^2 \theta^{\ast}) + 4 m_{\mu}^2(1 - \cos^2 \theta^{\ast})   - 4\sqrt{s \left( \frac{s}{4} - m_{\mu}^2 \right)}\cos \theta^{\ast} \right]\,.
\eeq
Substituting this into a numerator in Eq.~\eqref{eq:AFBdef} with the denominator from Eq.~\eqref{eq:AFBdenom},
the ALP contribution to the FB asymmetry is derived as
\beq
A_{\rm FB}^{\rm ALP} \simeq 
 \frac{3}{8 e^{4} }
\left( \int_{0}^1 d (\cos \theta^{\ast})\,  {\cal I}  - \int_{-1}^0 d( \cos \theta^{\ast})\,  {\cal I} \right)    \,.
\eeq
for $s \gg m_{\mu}^2$.
Note that the differential cross section is given as
\beq
\left(\frac{1}{2}\right)^2 \left| \mathcal{M}\left( e^+ e^- \to \mu^+ \mu^- \right)\right|^2 &= 
2 e^4 \frac{s^2+ 2 s k^2 + 2 k^4- 4 m_{\mu}^2 k^2 + 2 m_{\mu}^4 }{s^2}\non
&\phantom{=}
+ 2 e^2
\left[
\left|\left(y_V\right)_{e \mu} \right|^2 + \left|\left(y_A\right)_{e \mu}\right|^2   \right]
\frac{ k^4 +  m_{\mu}^2 \left(s - 2 k^2\right)+m_{\mu}^4}{s (k^2 - m_a^2)}  \non
&\phantom{=}
+ \left[ 
\left|\left(y_V\right)_{e \mu} \right|^2 + \left|\left(y_A\right)_{e \mu}\right|^2  \right]^2 
\left( \frac{k^2 - m_{\mu}^2}{k^2 - m_a^2} \right)^2\,,
\eeq
where the second line corresponds to the interference term.
 
\bibliography{ref}

\end{document}